\documentclass[british]{article}
\usepackage{mathpazo}
\usepackage[T1]{fontenc}
\usepackage[utf8]{inputenc}
\usepackage{geometry}
\usepackage[skip=\medskipamount]{parskip}
\usepackage{babel}
\usepackage{array}
\usepackage{mathtools}
\usepackage{bm}
\usepackage{amsmath}
\usepackage{amssymb}
\usepackage{cancel}
\usepackage{mathdots}
\usepackage{stmaryrd}
\usepackage{stackrel}
\usepackage{undertilde}
\usepackage{graphicx}
\PassOptionsToPackage{version=3}{mhchem}
\usepackage{mhchem}
\usepackage{esint}
\usepackage{microtype}
\usepackage[pdfusetitle,
 bookmarks=true,bookmarksnumbered=false,bookmarksopen=false,
 breaklinks=false,pdfborder={0 0 0},pdfborderstyle={},backref=false,colorlinks=false]
 {hyperref}

\makeatletter

\providecommand{\tabularnewline}{\\}

\numberwithin{equation}{section}

\@ifundefined{date}{}{\date{}}
\usepackage[margin=1cm]{caption}

\usepackage{tocloft}
\cftsetindents{subsubsection}{4em}{4em}

\newcommand*{\diff}{\mathop{}\!\mathrm{d}}
\DeclareMathOperator\supp{supp}

\usepackage{tensor}

\usepackage{fontawesome5}

\makeatother

\begin{document}
\title{Warp Drives and Closed Timelike Curves}
\author{\textbf{Barak Shoshany}{\small\thanks{Author to whom any correspondence should be addressed.}}\\
{\small\faIcon{envelope} \href{mailto:bshoshany@brocku.ca}{bshoshany@brocku.ca}$\qquad$\faIcon{orcid}
\href{https://orcid.org/0000-0003-2222-127X}{0000-0003-2222-127X}$\qquad$\faIcon{globe}
\href{https://baraksh.com/}{https://baraksh.com/}}\\
{\small\faIcon{university} \href{https://brocku.ca/}{Department of Physics, Brock University}}\\
{\small\faIcon{map-marker-alt} \href{https://goo.gl/maps/qscBMigohESxxczM7}{1812 Sir Isaac Brock Way, St. Catharines, Ontario, L2S 3A1, Canada}}\\
{\small}\\
\textbf{Ben Snodgrass}\\
{\small\faIcon{envelope} \href{mailto:benedict.snodgrass@math.uni-freiburg.de}{benedict.snodgrass@math.uni-freiburg.de}$\qquad$\faIcon{orcid}
\href{https://orcid.org/0009-0002-0546-5045}{0009-0002-0546-5045}}\\
{\small\faIcon{university} \href{https://www.math.uni-freiburg.de/institut/abteilungen/rein.html?l=en}{Department of Pure Mathematics, University of Freiburg}}\\
{\small\faIcon{map-marker-alt} \href{https://maps.app.goo.gl/zhay5Wh7jXrw6Gvi6}{Ernst-Zermelo-Straße 1, 79104 Freiburg im Breisgau, Germany}}}
\maketitle
\begin{abstract}
It is commonly accepted that superluminal travel may be used to facilitate
time travel. This is a purely special-relativistic argument, using
the fact that for observers in two frames of reference, separated
by a spacelike interval, the non-causal (spacelike) future of one
observer includes part of the causal past of the other. In this paper
we provide a concrete realization of this argument in a curved general-relativistic
spacetime, using warp drives as the means of faster-than-light travel.
By generalizing the usual warp drive metric to allow for a non-unit
lapse function, we allow the warp drive to switch between reference
frames in a purely geometric way. With an additional modification
allowing the warp drive to have compact support, this permits us to
glue two warp drives together to construct a closed timelike geodesic,
such that a test particle following the geodesics of the two warp
drives travels back to its own past. This provides a precise mathematical
model for the connection between faster-than-light travel and time
travel in general relativity, and the first such model to be explicitly
formulated using two warp drives. We also give a detailed discussion
of weak energy condition violations in the non-unit-lapse warp drive.
\end{abstract}
\tableofcontents{}

\section{Introduction}\label{Sec:=000020Intro}

\subsection{Warp drives}

Since Alcubierre's seminal 1994 paper \cite{Alcubierre_1994}, the
warp drive has developed from a vague, science-fictional concept to
a subject of genuine scientific interest \cite{Shoshany_2019,Lobo,Krasnikov}.
Whilst not generally considered to be a realistic option for human
interstellar travel in the near future, it serves as an interesting
theoretical model, opening up new insights into the surprising possibilities
that arise from the curved spacetime geometry of general relativity.
Studying warp drives may also expose pathologies in the theory itself.

The Alcubierre warp drive is defined by the metric 
\begin{equation}
\diff s^{2}=-\diff t^{2}+\diff x^{2}+\diff y^{2}+(\diff z-fv\diff t)^{2},\label{Alcubierre}
\end{equation}
where $f\equiv f(x,y,z-\zeta(t))$ with $f(0,0,0)=1$, $f\rightarrow0$
at spatial infinity, and $v\equiv v(t)\equiv\partial_{t}\zeta(t)$.
Note that this metric is flat on a hypersurface $t=\text{constant}$
if and only if $v(t)=0$. It can easily be shown that the path 
\begin{equation}
\alpha(\tau)=(\tau,0,0,\zeta(\tau)),\qquad\tau\in(-\infty,\infty),
\end{equation}
is a geodesic, parameterised by proper time $\tau$.

The idea behind the warp drive is to take advantage of a loophole
in general relativity: that while all massive objects are constrained
to move along timelike paths, space itself has no such restriction.
Roughly speaking, a warp drives exploits this by having a shell of
curved spacetime (the `warp bubble') embedded in a flat background
spacetime, which can accelerate its flat interior to arbitrarily high
speeds, without the passengers inside feeling any acceleration whatsoever.
An observer inside the warp bubble will follow a timelike geodesic,
but its speed is unbounded – and in particular, it is not limited
by the speed of light, so the observer can effectively travel between
two spacelike-separated points.

There are numerous problems with warp drives as they have so far been
conceived, even setting aside the extreme engineering difficulties.
The most notable are their seemingly-inevitable violation of the energy
conditions \cite{Curiel:2014zba,Santiago_2022}, the typically vast
amounts of (negative) energy required to sustain them \cite{Alcubierre_1994,Broeck_1999},
and the formation of event horizons \cite{Alcubierre_1994,Gonz_lez_D_az_2000}.

Despite those issues, warp drives are an important theoretical tool
in general relativity, most importantly in the study of causality
and its violations. In this paper we will focus on the question of
how warp drives can be used to create closed timelike curves. For
this purpose we will propose a generalised warp drive model, and we
will spend some time studying one of the above issues, the violation
of the weak energy condition, in the context of this new model.

\subsection{Superluminal travel and time travel}

In this paper, we shall implement the following well-known method
of using superluminal travel to facilitate time travel. Consider two
reference frames, $S$ and $S'$, with some relative velocity $u$
along the $x$ axis. Suppose one can travel a large distance at a
superluminal speed\footnote{Throughout this paper we will be using units where $c=1$.}
$v>1$ in a given reference frame. Then, if $u$ is large enough for
a given $v$ (as discussed in detail in Section \ref{Sec:=000020CTC=000020req}),
one can travel back in time by travelling superluminally between two
points in the $S$ frame, transitioning to the $S'$ frame, and returning
superluminally in the $S'$ frame to the starting point.

\begin{figure}[!h]
\centering
\centering{}\includegraphics[width=1\textwidth]{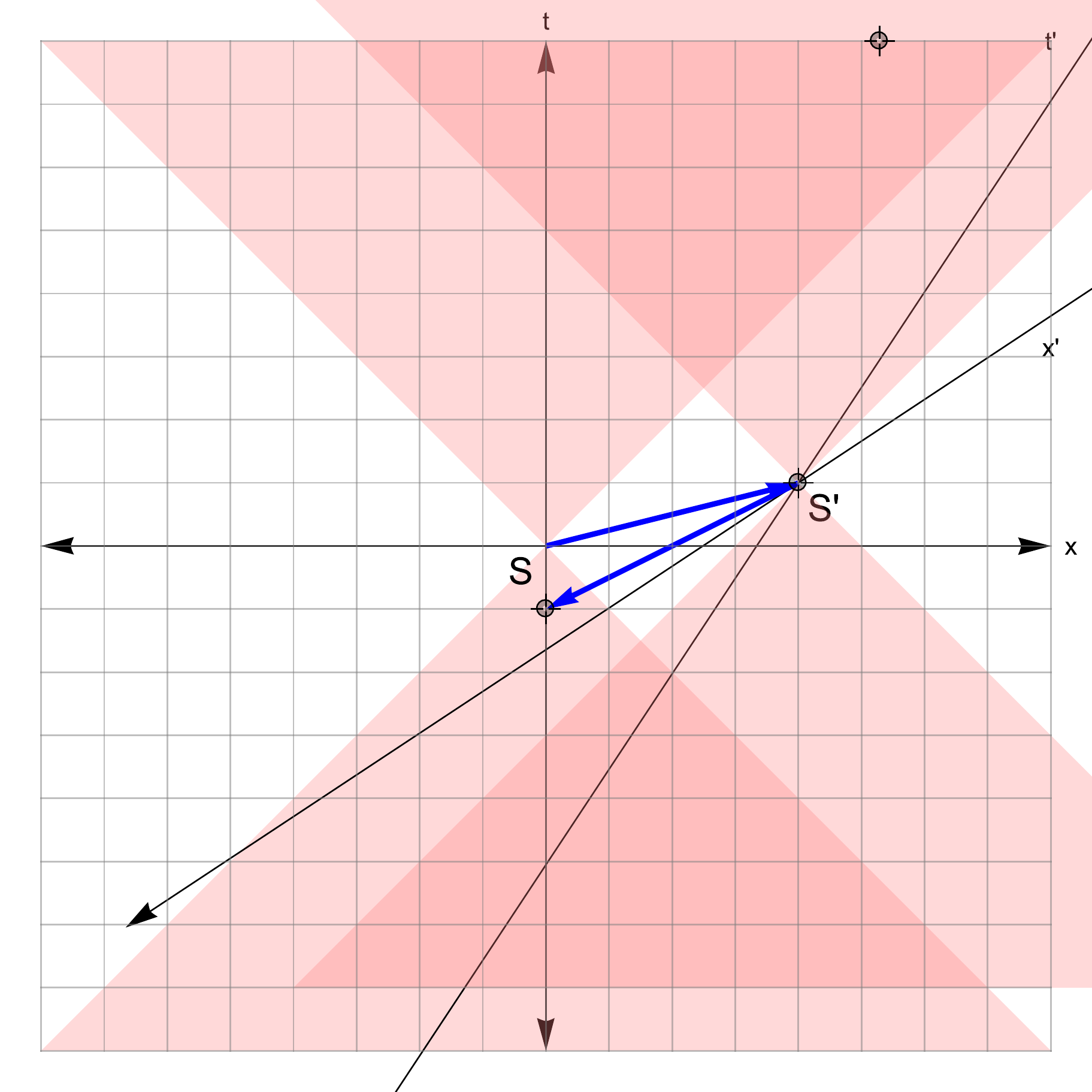} \caption{Spacetime diagram showing how superluminal travel may allow for time
travel. In the unprimed coordinates (representing $S$), the path
moving in the positive $x$ direction is moving forwards in $t$,
and in the primed coordinates (representing $S'$), the path moving
along the negative $x$ direction is moving forwards in $t'$. However,
each path viewed in the other frame is travelling backwards in their
respective time coordinates. Here the relative velocity between the
frames is $u\approx0.66$ (Lorentz factor $\gamma\approx1.33$), and
the speed of the warp drive in both directions is $v\approx4$.}
\label{Fig:=000020CTC}
\end{figure}

This is proven mathematically in Section \ref{Sec:=000020CTC=000020req},
but it is visually clear from the spacetime diagram\footnote{The reader may reproduce this spacetime diagram interactively using
the Mathematica notebook available at \href{https://github.com/bshoshany/spacetime-diagrams}{https://github.com/bshoshany/spacetime-diagrams}.} in Figure \ref{Fig:=000020CTC}. Here, the traveller uses two warp
drives to follow the path of the blue arrows, which both represent
superluminal paths in $S$ and $S'$ respectively. First, they travel
from $x=0$ out along the upper arrow, finishing at a speed $u$ and
perhaps landing on a space station at rest in $S'$. Then, they travel
back along the lower arrow\footnote{In this diagram, changes in velocity happen instantaneously. This
could of course be smoothed out to make the acceleration finite everywhere.}. The traveller arrives back at $x=0$ \emph{earlier} than when they
left, so they have travelled backwards in time.

The light cones of $S$ and $S'$ are highlighted in pink; of course,
the blue paths are outside the light cones, as they are spacelike.
Note that along the upper arrow, moving in the positive $x$ direction,
the traveller is moving forwards in time according to $S$, but backwards
in time according to $S'$. Similarly, along the lower arrow, moving
in the negative $x$ direction, the traveller is moving forwards in
time according to $S'$, but backwards in time according to $S$.
This is how the `trick' works: for any object moving superluminally
in one frame of reference, there is another frame of reference in
which they are moving backwards in time. 

\subsection{Warp drives and closed timelike curves}

To our knowledge, \cite{Everett_1996} was the first to argue that
warp drives can be used to construct closed timelike curves in the
fashion described above; see also a summary in \cite{Lobo_2010}.
While this work serves as a clear and useful proof of principle, it
has several shortcomings, which we would like to highlight – and resolve
– in this paper.

First, as the traveller moves from $S$ to $S'$, they must also change
their reference frame. The reason is that if they stayed in the reference
frame of $S$, they must still move towards the future of $S$ (i.e.
towards increasing values of $t$), so time travel would be impossible.

There is something seemingly `magical' about the frame transition
from $S$ to $S'$; before the transition, even if the traveller moved
at arbitrarily high speeds, $v\to\infty$, they still could never
move back in time. However, once they land on the space station at
rest in $S'$, and a Lorentz transformation is performed, travel to
the past suddenly becomes possible.

In this paper, we will provide a concrete warp drive metric that takes
this `landing' into account, and therefore makes this `magical'
part precise. One issue that we will have to overcome is that warp
drives, as previously formulated, reside fully within a single reference
frame from beginning to end, and thus do not allow such rest frame
transitions to occur. Of course, this is not an issue if one is travelling
within the warp bubble in a spaceship; they can simply accelerate
using the spaceship's rockets to facilitate the landing on the moving
target. However, this does introduce a significant conceptual hurdle,
namely that it is impossible to create a closed timelike \emph{geodesic}
using a warp drive alone.

A closed timelike \emph{curve} can be constructed, for example with
rocket propulsion or perhaps some sort of conveyor belt, but it will
not be a \emph{geodesic} – meaning that it is not a feature of the
spacetime itself, but requires external forces to work. Furthermore,
including external acceleration will considerably complicate the theoretical
and mathematical analysis; we cannot simply write down a metric that
will contain a closed timelike curve, we would have to consider the
external acceleration separately. To resolve this issue, we will generalise
the warp drive metric such that it is inherently capable of changing
frames as needed.

The second issue is that the usual warp drives are non-compact, meaning
that the two warp drives – one from $S$ to $S'$, another from $S'$
back to $S$ – overlap. In previous work, it was simply assumed that
this overlapping was negligible, leaving the geodesics inside the
two warp drives unaffected. In this paper we will make this construction
more precise and explicit by introducing \emph{compact} warp drives,
which do not overlap with each other.

The third and final issue is that, in previous work, the metric of
one warp drive was considered, but a complete metric containing both
the outgoing and incoming warp drives, as well as an \emph{explicit}
closed timelike geodesic, was not derived. Our main result in this
paper will be a complete and explicit spacetime metric, (\ref{CTC=000020metric}),
containing two warp drives of a new, generalised kind, which:
\begin{itemize}
\item Begin and end in different rest frames,
\item Do not overlap with each other, and
\item Create an explicit closed timelike geodesic.
\end{itemize}
This construction will provide a much more solid foundation to the
widely-held belief that faster-than-light travel allows for time travel.
Furthermore, the explicit metric we will construct will provide a
foundation for future investigations of various consequences of time
travel, such as time travel paradoxes \cite{Shoshany_2019,ShoshanyHauser,ShoshanyWogan,ShoshanyStober},
in a concrete general-relativistic setting.

Our paper relies on the definition of a new generalised warp drive
with non-unit lapse function. Therefore, it would be interesting to
investigate whether this resolves what is often considered to be the
most problematic feature of warp drives, namely the violation of the
pointwise weak energy condition (WEC). We do this by extending the
arguments presented in \cite{Santiago_2022} to allow for a non-unit
lapse.

Unfortunately, we find that this does not help to avoid WEC violations,
subject to some reasonable conditions. We also give an analysis of
the case of zero Eulerian energy density and a curl-free shift vector
field, as this was not accounted for in \cite{Santiago_2022}. Interestingly,
we conclude that the WEC violations are likely nothing more than an
artefact of the particular class of metrics, and have nothing to do
with the metrics' potential to describe warp drives, or superluminal
travel in general.

This paper introduces many different mathematical symbols. To prevent
confusion and help make the mathematical formulas presented in the
paper more clear, we provide a convenient table of symbols in Appendix
\ref{sec:Tables-of-symbols}. We denote 4-dimensional spacetime indices
using Greek letters, and 3-dimensional spatial indices using lowercase
English letters.

Throughout the paper, we have used both the Mathematica package OGRe
(An Object-Oriented General Relativity Package) \cite{Shoshany2021_OGRe}
and a beta version of the Python package OGRePy \cite{Shoshany2024_OGRePy}
to facilitate calculations of curvature tensors, geodesics, and other
relevant geometrical quantities.

\section{Geodesic rest frame transitions}\label{Sec:=000020Natario=000020failure}

\subsection{Warp drives with unit lapse}

First, we give a preliminary definition of a \emph{geodesic rest frame
transition}, which will suffice for our purposes. We give a more general
definition and thorough analysis in Appendix \ref{Sec:=000020FT=000020general=000020analysis},
from a coordinate-independent starting point.

We take a general metric $g_{\mu\nu}$ expressed in the ADM formalism
and coordinates $(t,x,y,z)$ as\footnote{In this paper, as in most warp drive papers, we define the shift vector
$\bm{\beta}$ as the negative of what is usually used in other areas.
With this convention, passengers in the warp drive (usually) move
with a 3-velocity of $\bm{\beta}$.} 
\begin{equation}
\diff s^{2}=-N^{2}\diff t^{2}+\gamma_{ij}(\diff x^{i}-\beta^{i}\diff t)(\diff x^{j}-\beta^{j}\diff t).\label{gen=000020ADM}
\end{equation}
The metric is assumed to be asymptotically flat in these coordinates,
that is, as $r\equiv\sqrt{x^{2}+y^{2}+z^{2}}\rightarrow\infty$, 
\[
N\rightarrow1,\qquad\bm{\beta}\rightarrow\mathbf{0},\qquad\gamma_{ij}\rightarrow\delta_{ij}.
\]
We say that (\ref{gen=000020ADM}) allows for a \emph{geodesic rest
frame transition} if: 
\begin{enumerate}
\item $\exists T_{flat}>0$ such that if $t\in(-\infty,0]\cup[T_{flat},\infty)$,
$g_{\mu\nu}=\eta_{\mu\nu}$, where $\eta_{\mu\nu}$ is the Minkowski
metric.\footnote{This may seem unnecessarily restrictive. Could a free-falling observer
inside a warp drive not be cast out of the warp drive with a non-zero
velocity, without having the warp drive completely vanish? Yes, in
principle, but we argue that once they have left the warp drive, for
their 4-velocity to be well-defined with respect to the original frame,
they must be in a simply connected region of flat spacetime which
stretches to future timelike infinity. If this is the case, then there
will be no problem in `flattening' the warp drive, since this cannot
affect the 4-velocity of the observer any more. Therefore, any metric
describing a rest-frame-transitioning warp drive could be easily modified
to fit with this definition, without changing the environment of $\Gamma_{RFT}$
at all. For a detailed discussion, see Appendix \ref{Sec:=000020FT=000020general=000020analysis}.} 
\item There exists a timelike geodesic $\Gamma_{RFT}$ such that for $t\leq0$,
$\Gamma_{RFT}$ has tangent vector $v^{\mu}=(1,\mathbf{0})$, \emph{and}
that for $t\geq T_{flat}$, $\Gamma_{RFT}$ has tangent vector $v^{\mu}\neq(1,\mathbf{0}).$\footnote{Since the sections of $\Gamma_{RFT}$ with $t\notin(0,T_{flat})$
are in flat spacetime, these tangent vectors have constant components
in these regions.} 
\end{enumerate}
For $t\notin(0,T_{flat})$, the coordinates $(t,x,y,z)$ describe
an inertial reference frame, since $g_{\mu\nu}=\eta_{\mu\nu}$. This
encapsulates the idea of an observer starting at rest in this reference
frame and finishing at rest in a different reference frame. 

We shall now demonstrate that warp drives do not allow for rest frame
transitions. We will focus specifically on warp drives of the Natário
class \cite{Nat_rio_2002}\footnote{This class includes the original Alcubierre drive, and should not
be confused with the zero-expansion warp drives, described in the
second half of Natário's paper.} , which is the set of asymptotically-flat spacetimes of the form
(\ref{gen=000020ADM}) with $N=1$ and $\gamma_{ij}=\delta_{ij}$:
\begin{equation}
\diff s^{2}=-\diff t^{2}+\delta_{ij}(\diff x^{i}-\beta^{i}\diff t)(\diff x^{j}-\beta^{j}\diff t).\label{Natario}
\end{equation}
Define $\Sigma_{t^{*}}\equiv t^{-1}(t^{*})$ for $t^{*}\in\mathbb{R}$.
The normal vector field to $\Sigma_{t}$ is given by 
\begin{equation}
\begin{gathered}\mathbf{n}\equiv-\diff t\\
\implies n_{\mu}=(-1,\mathbf{0}),\qquad n^{\mu}=(1,\bm{\beta}).
\end{gathered}
\label{normal=000020N1}
\end{equation}
Consider an observer starting at rest in the rest frame described
by these coordinates at a time $t<0$, with 4-velocity $(1,0,0,0)$
(as they would along $\Gamma_{RFT}$, if it exists). This means that
the observer is initially Eulerian, by which we mean that their 4-velocity
coincides with the normal vector field. Consider the following general
formula, valid in any ADM spacetime:
\begin{equation}
n^{\nu}\nabla_{\nu}n_{\mu}=\frac{1}{N}(\nabla_{\mu}N+n_{\mu}\mathcal{L}_{\mathbf{n}}N),\label{Eulerian=000020acceleration}
\end{equation}
where $\mathcal{L}$ denotes the Lie derivative. From this formula,
one can easily show that the integral curves of the vector field $n^{\mu}$
(the paths of Eulerian observers) are geodesics if and only if $\partial_{i}N=0\iff N=N(t)$.
This is clearly the case for (\ref{Natario}), so the observer, wherever
they go, must have 4-velocity $n^{\mu}$. 

From the known expression for an inverse metric expressed in the ADM
formalism, we can see that 
\begin{equation}
n^{\mu}=-Ng^{0\mu}=-g^{0\mu},
\end{equation}
since, as $t\rightarrow T_{flat}$, $g^{0\mu}\rightarrow\eta^{0\mu}=(-1,0,0,0)$
and $n^{\mu}\rightarrow(1,\mathbf{0})$. The observer must have this
same 4-velocity, and we see that they have returned to being at rest
in the original frame of reference. Therefore, (\ref{Natario}) does
\emph{not} allow for rest frame transitions, so it cannot be used
to facilitate time travel using the method described in the introduction
without complicating the model with an external, non-geodesic source
of acceleration. 

The above argument can easily be extended to the case where $N=N(t)$,
$\delta_{ij}\rightarrow\gamma_{ij}$ for arbitrary $N(t)>0$ and positive-definite
$\gamma_{ij}$, since Eulerian observers still follow geodesics and
we must still have $n^{\mu}\rightarrow(1,\mathbf{0})$ as $t\rightarrow T_{flat}$.
Therefore, in order to find a metric allowing for a rest frame transition,
we have no choice but to introduce spatial dependence into the lapse.
This will be the subject of the next section.

\subsection{Warp drives with non-unit lapse}\label{Sec:=000020FT=000020warp=000020drive}

In this section, we introduce a generalised warp drive that starts
from rest in some reference frame and travels along an arbitrary path
$\mathbf{r}(t)$ for $0<t<T_{1}$ where $T_{1}>0$, finishing at a
constant 3-velocity $(0,0,u)$ for $t\geq T_{1}$. We then show how
we can manipulate the metric so that the warp curvature flattens whilst
keeping the free-falling passengers moving at a speed $u$ along the
$z$ axis. This means that we have found a warp drive capable of `landing'
on a target moving at a speed $u$.

First, we introduce a generic warp drive following an arbitrary path
$\mathbf{r}(t)=(X(t),Y(t),Z(t))$, with a shift vector $\bm{\beta}=a(t)\mathbf{v}(t,\mathbf{x})$.
We set $\partial_{t}\mathbf{r}(t)=\mathbf{v}(t,\mathbf{r}(t))$ so
that $(t,\mathbf{r}(t))$ is an integral curve of $(1,\mathbf{v}(t,\mathbf{x}))$\footnote{In the original Alcubierre drive (\ref{Alcubierre}), this condition
is a generalisation of setting $f$ as a function of $z-\zeta(t)$
and then setting the metric coefficient as $vf$ where $v(t)=\partial_{t}\zeta(t)$
and $f=1$ at the warp drive centre.}. Our new metric is given by 
\begin{equation}
\diff s^{2}=-N^{2}\diff t^{2}+\delta_{ij}(\diff x^{i}-\beta^{i}\diff t)(\diff x^{j}-\beta^{j}\diff t).\label{Natario=000020mod}
\end{equation}
The introduction of the function $a$ and a non-unit lapse $N$ are
our modifications to the Natário form of the metric, and their significance
shall soon become clear. We also set $\mathbf{v}$ such that 
\begin{equation}
(\partial_{i}v^{j})(t,\mathbf{r}(t))=0,
\end{equation}
that is, $\mathbf{v}$ is flat at the warp drive centre $(t,\mathbf{r}(t))$,
and as usual, we also require $\mathbf{v}\rightarrow\mathbf{0}$ at
spatial infinity. All functions describing the metric are assumed
to be at least $C^{2}$, such that the Riemann tensor is continuous. 

The geodesic Lagrangian is 
\begin{equation}
\mathcal{L}_{g}=\frac{1}{2}\left(-N^{2}\Dot{t}^{2}+\delta_{ij}(\Dot{x}^{i}-\beta^{i}\Dot{t})(\Dot{x}^{j}-\beta^{j}\Dot{t})\right),
\end{equation}
where we shall use a dot to denote differentiation with respect to
proper time $\tau$. The Euler-Lagrange equations then imply 
\begin{equation}
\begin{gathered}0=\frac{\diff}{\diff\tau}(-N^{2}\Dot{t}-\delta_{jk}\beta^{j}(\Dot{x}^{k}-\beta^{k}\Dot{t}))+\Dot{t}^{2}N\partial_{t}N+\Dot{t}\cdot\delta_{jk}(\partial_{t}\beta^{j})(\Dot{x}^{k}-\beta^{k}\Dot{t}),\\
0=\frac{\diff}{\diff\tau}(\Dot{x}^{i}-\beta^{i}\Dot{t})+\Dot{t}^{2}N\partial_{x^{i}}N+\Dot{t}\cdot\delta_{jk}(\partial_{x^{i}}\beta^{j})(\Dot{x}^{k}-\beta^{k}\Dot{t}).
\end{gathered}
\label{geo=000020eqns}
\end{equation}
The unit normal vector field to hypersurfaces of constant $t$, $\Sigma_{t}$,
is given by 
\begin{gather}
\mathbf{n}\equiv-N\diff t\quad\implies\quad n_{\mu}=(-N,\mathbf{0}),\qquad n^{\mu}=\frac{1}{N}(1,\bm{\beta}).
\end{gather}
Again, an integral curve of this vector field will satisfy the geodesic
equations if and only if $\partial_{i}N=\mathbf{0}$, as can be seen
from (\ref{geo=000020eqns}). The path $(\tau,\mathbf{r}(\tau))$
is thus a geodesic wherever $N=1$ and $a=1$, as it is an integral
curve of $n^{\mu}$. It is also parameterised by proper time. 

Now take $0<T_{1}<T_{2}$ and consider the case that 
\begin{equation}
\begin{cases}
\mathbf{v}(t,\mathbf{x})=\mathbf{0} & t\leq0,\ \mathbf{x}\in\mathbb{R}^{3},\\
\mathbf{v}(t,\mathbf{r}(t))=(0,0,u) & t\geq T_{1},
\end{cases}\label{v}
\end{equation}
for some fixed $0<u<1$, and $\mathbf{v}$ is unconstrained for $0<t<T_{1}$.
We take $a(t)$ as follows: 
\begin{equation}
\begin{cases}
a(t)=1 & t\leq T_{1},\\
0<a(t)<1 & T_{1}<t<T_{2},\\
a(t)=0 & t\geq T_{2},
\end{cases}
\end{equation}
and set the lapse as 
\begin{equation}
N=1-b\cdot(z-Z)s.\label{lapse}
\end{equation}
where $s\rightarrow0$ at spatial infinity with $s(t,\mathbf{r}(t))=1$,
$(\partial_{i}s)(t,\mathbf{r}(t))=0$ and 
\begin{equation}
\begin{gathered}b(t)\equiv-u\lambda^{2}\cdot\partial_{t}a,\\
\lambda\equiv(1-u^{2}(1-a)^{2})^{-\frac{1}{2}}.
\end{gathered}
\label{b=000020from=000020a}
\end{equation}
We call $a$ and $b$ the \emph{transition functions} for the shift
vector and lapse, respectively. $s$ and $a$ must also be chosen
such that $N>0$ everywhere. 

Note that for $t\geq T_{2}$, $a$ and $b$ vanish, so $g_{\mu\nu}=\eta_{\mu\nu}$,
and the metric becomes Minkowskian (so $T_{2}$ corresponds to $T_{flat}$
in the previous section). $b(t)$ vanishes for $t\notin(T_{1},T_{2})$,
so $N=1$ for $t\notin(T_{1},T_{2})$, and Eulerian observers follow
geodesics here, so taking 
\begin{equation}
\chi_{1}(\tau)\equiv(\tau,\mathbf{r}(\tau)),\qquad0\leq\tau\leq\tau_{1}\equiv T_{1},
\end{equation}
we see $\chi_{1}([0,\tau_{1}])$ is a geodesic. However, in the region
with $t\in(T_{1},T_{2})$, Eulerian observers do not follow geodesics.
For $T_{1}\leq t\leq T_{2}$, consider the path $\chi_{2}$ satisfying
\begin{equation}
\begin{gathered}\Dot{\chi_{2}}(\tau)=(\lambda,0,0,\lambda u),\\
\chi_{2}(\tau_{1})=(T_{1},\mathbf{r}(T_{1})),
\end{gathered}
\label{ini=000020cond=0000201}
\end{equation}
so that at $\tau=\tau_{1}>0$, $a$ may start to change. Since $0\leq a\leq1$
and $|u|<1$, we have
\begin{equation}
1\leq\lambda\leq\gamma_{u},
\end{equation}
where 
\[
\gamma_{u}\equiv(1-u^{2})^{-\frac{1}{2}}
\]
is the Lorentz factor, and this ordinary differential equation has
a unique, smooth solution for all $\tau\geq\tau_{1}$\footnote{It can also be seen that since $\frac{\diff t}{\diff\tau}=\lambda\geq1$
is bounded between two finite values, the duration of the transition
as measured by proper time is bounded and strictly less than the duration
as measured by coordinate time.}. We now demonstrate that the solution to (\ref{ini=000020cond=0000201})
is a geodesic. 

For $\chi_{2}$, 
\begin{equation}
\frac{\diff z}{\diff t}=\frac{\diff z}{\diff\tau}\cdot\frac{\diff\tau}{\diff t}=\lambda u\cdot\frac{1}{\lambda}=u,
\end{equation}
so parameterised by coordinate time, this is the path 
\begin{equation}
(\chi_{2}\circ\tau)(t)=(t,X(T_{1}),Y(T_{1}),Z(T_{1})+u(t-T_{1}))=(t,\mathbf{r}(t)),
\end{equation}
and looking at (\ref{lapse}), $N=1$ along this path (but crucially,
$\partial_{z}N\neq0$). As can be checked from (\ref{Natario=000020mod}),
this has the consequence that $\Dot{\chi}_{2}$ is normalised, i.e.
$g(\Dot{\chi}_{2},\Dot{\chi}_{2})=-1$, and we see that $\tau$ is
the proper time of the path, with 
\begin{equation}
\tau(t)=T_{1}+\int_{T_{1}}^{t}\sqrt{-g(\partial_{t}\chi_{2},\partial_{t}\chi_{2})}\diff t',\qquad T_{1}<t\leq T_{2}.
\end{equation}
Taking the first geodesic equation, along $\chi_{2}$, we have 
\begin{equation}
\begin{aligned}0 & =\frac{\diff}{\diff\tau}(-1^{2}\cdot\lambda-\lambda\cdot(u-au)\cdot au)+\lambda^{2}\cdot1\cdot ub+u\lambda\partial_{t}a(\lambda u(1-a))\\
 & =-\frac{\diff}{\diff\tau}(\lambda+\lambda u^{2}a(1-a))+\lambda^{2}ub+\lambda u^{2}\frac{\diff a}{\diff\tau}(1-a)\\
 & =-\Dot{\lambda}+\lambda^{2}ub-au^{2}\frac{\diff}{\diff\tau}(\lambda(1-a))\\
 & =-\Dot{\lambda}-u^{2}a(1-a)\Dot{\lambda}+u^{2}\lambda a\Dot{a}+\lambda^{2}ub\\
 & =u^{2}\lambda\Dot{a}(\lambda^{2}(1-a)+au^{2}\lambda^{2}(1-a)^{2}+au^{2})+\lambda^{2}ub\\
 & =u^{2}\lambda^{3}\Dot{a}((1-a)+au^{2}(1-a)^{2}+a(1-u^{2}(1-a)^{2}))+\lambda^{2}ub\\
 & =u^{2}\lambda^{3}\Dot{a}+\lambda^{2}ub\\
\implies & 0=u\lambda^{2}\partial_{t}a+b,
\end{aligned}
\label{geo=000020simp}
\end{equation}
where in the fifth line we have used the identity $\Dot{\lambda}=-\lambda^{3}u^{2}(1-a)\Dot{a}$,
and in the sixth the definition of $\lambda$. Therefore, we see that
we recover (\ref{b=000020from=000020a}) as a necessary condition
for $\chi_{2}$ to be geodesic. 

The geodesic equations for $x(\tau)$ and $y(\tau)$ are trivially
satisfied since, along $\chi_{2}$, 
\[
\Dot{x}=\beta^{x}=\partial_{x}N=\Dot{y}=\beta^{y}=\partial_{y}N=0.
\]
So finally, considering the equation for $z(\tau)$: 
\begin{equation}
\begin{aligned}0 & =\frac{\diff}{\diff\tau}(\lambda u(1-a))-\lambda^{2}\cdot1\cdot b+0\\
 & =\frac{\diff}{\diff\tau}(\lambda u(1-a))-\lambda^{2}b,
\end{aligned}
\end{equation}
where we have used that $(\partial_{i}\beta^{j})(t,\mathbf{r}(t))=0\implies\partial_{z}\beta^{z}=0$
and $(\partial_{i}s)(t,\mathbf{r}(t))=0$ along $\chi_{2}$. Now looking
at the third line in (\ref{geo=000020simp}), we see this can be written
as 
\begin{equation}
\begin{aligned} & 0=\frac{1}{au}(-\Dot{\lambda}+\lambda^{2}ub)-\lambda^{2}b\\
\implies & 0=u\lambda^{2}\partial_{t}a+b.
\end{aligned}
\end{equation}
Therefore, (\ref{b=000020from=000020a}) is a necessary \emph{and}
sufficient condition for $\chi_{2}$ to be geodesic.

Finally, we note that the combination of the two above paths 
\begin{equation}
\chi(\tau)\equiv\begin{cases}
\chi_{1}(\tau) & 0\leq\tau<\tau_{1},\\
\chi_{2}(\tau) & \tau_{1}\leq\tau\leq\tau_{2},
\end{cases}
\end{equation}
where $\tau_{2}\equiv\tau(T_{2})$, is also a geodesic. This can be
seen by noting that both the position and tangent vector of $\chi_{1}$
at $\tau=\tau_{1}$ match those of $\chi_{2}$ at $\tau=\tau_{1}$,
so extending the geodesic $\chi_{1}$ is equivalent to solving the
geodesic equation with initial conditions 
\begin{equation}
\begin{gathered}\chi(\tau_{1})=(T_{1},\mathbf{r}(T_{1})),\\
\Dot{\chi}(\tau_{1})=(1,0,0,u),
\end{gathered}
\end{equation}
which is exactly $\chi_{2}$. Written another way, the entire curve
\begin{equation}
\Gamma_{RFT}\equiv\{p\in M:\exists\,t\in[0,T_{2}]:p=(t,\mathbf{r}(t))\}\label{Gamma=000020defn}
\end{equation}
is a geodesic, and it satisfies the requirements described in the
previous section. This means that a free-falling observer starting
at rest in the centre of the bubble does not fall out of it, and their
final 4-velocity at time $t=T_{2}$ is $(\gamma_{u},0,0,\gamma_{u}u)$
– not the same as their initial 4-velocity of $(1,0,0,0)$. Since
for $t\geq T_{2}$ we have $g_{\mu\nu}=\eta_{\mu\nu}$, we see that
an observer following this geodesic has transitioned between rest
reference frames.

\section{Creating a closed timelike geodesic }\label{Sec:=000020Making=000020CTC}

\subsection{The double-warp-drive metric}\label{Sec:=000020CTC=000020gen=000020construct}

In this section, we demonstrate how the above method of performing
a rest frame transition may be employed to write down a metric describing
a spacetime containing a \emph{closed timelike geodesic}. To the authors'
knowledge, this is the first complete and well-defined example of
a spacetime metric which \emph{explicitly} uses warp drives to create
a closed timelike curve, and furthermore, the closed timelike curve
we find is a geodesic, requiring no acceleration due to external forces,
which would complicate the theoretical and mathematical analysis.

The idea, as described in the introduction, is to have an observer
start at rest in a reference frame $S$, travel in a superluminal
warp drive of the form (\ref{Natario=000020mod}), transition to a
different reference frame\footnote{We use $\hat{S}$ here, instead of $S'$ as in the introduction, to
facilitate more concise notation of related parameters.} $\hat{S}$, and then travel back to the starting point in a similar
warp drive. It will turn out that if the two warp drives move fast
enough, this will result in the observer finishing the journey at
a point in the causal past of their departure. 

First, we introduce a generic spacetime, which we suggestively call
$M_{CTC}$, containing two warp drives that take an observer on a
journey to and from rest at a given spatial point in $S$, which we
shall take to be $(x,y,z)=(0,0,0)$. For the return journey, we shall
need to define a warp drive in the Lorentz-boosted frame $\hat{S}$,
which moves at speed $u$ along the $z$ axis with respect to $S$.
We thus take the metric 
\begin{equation}
\diff\hat{s}^{2}=-\hat{N}^{2}\diff\hat{t}^{2}+\delta_{\hat{i}\hat{j}}(\diff\hat{x}^{\hat{i}}-\hat{\beta}^{\hat{i}}\diff\hat{t})(\diff\hat{x}^{\hat{j}}-\hat{\beta}^{\hat{j}}\diff\hat{t}),\label{Natario=000020mod=000020hat}
\end{equation}
where $(\hat{t},\hat{x},\hat{y},\hat{z})$ is the Lorentz-boosted
coordinate system, using the convention that indices with hats denote
$\hat{S}$ coordinates and those without hats denote $S$ coordinates.
The functions $\hat{N}$, $\hat{\bm{\beta}}$, $\hat{\mathbf{v}}$,
$\hat{a}$, $\hat{b}$, $\hat{s}$, $\hat{\mathbf{r}}$, $\hat{X}$,
$\hat{Y}$, and $\hat{Z}$ are defined exactly analogously to their
$S$ counterparts, but with $\hat{S}$ coordinates as their arguments.
There are only two other changes: 
\begin{enumerate}
\item The critical times, $t=0,T_{1},T_{2}$, have now become $\hat{t}=\hat{T}_{0},\hat{T}_{1},\hat{T}_{2}$
respectively. 
\item $\hat{\mathbf{v}}$ is set such that the warp drive finishes at rest
in $S$: 
\end{enumerate}
\begin{equation}
\begin{cases}
\mathbf{\hat{v}}(\hat{t},\mathbf{\hat{x}})=\mathbf{0} & \hat{t}\leq\hat{T}_{0},\ \hat{\mathbf{x}}\in\mathbb{R}^{3},\\
\mathbf{\hat{v}}(\hat{t},\mathbf{\hat{r}}(\hat{t}))=(0,0,-u) & \hat{t}\geq\hat{T}_{1}.
\end{cases}\label{v=000020hat}
\end{equation}
Applying the Lorentz transformation 
\begin{equation}
\begin{gathered}\hat{t}=\gamma_{u}(t-uz),\\
\hat{x}=x,\\
\hat{y}=y,\\
\hat{z}=\gamma_{u}(z-ut),
\end{gathered}
\label{Lor=000020trans}
\end{equation}
gives the metric (\ref{Natario=000020mod=000020hat}) in $S$ coordinates
as 
\begin{equation}
\begin{gathered}\diff s^{2}=-\hat{N}^{2}\gamma_{u}^{2}(\diff t-u\diff z)^{2}+(\diff x-\hat{\beta}^{x}\gamma_{u}(\diff t-u\diff z))^{2}\\
+(\diff y-\hat{\beta}^{y}\gamma_{u}(\diff t-u\diff z))^{2}+\gamma_{u}^{2}((\diff z-u\diff t)-\hat{\beta}^{z}(\diff t-u\diff z))^{2}
\end{gathered}
\label{Natario=000020mod=000020hat=000020S}
\end{equation}
where all the functions have their arguments in terms of $S$ coordinates. 

Now we put these two warp drive spacetimes together. Defining $h_{\mu\nu}$
as the perturbation\footnote{If $\prescript{(1)}{}{g}_{\mu\nu}$ is the metric (\ref{Natario=000020mod}),
then $h_{\mu\nu}\equiv\prescript{(1)}{}{g}_{\mu\nu}-\eta_{\mu\nu}$.
Similarly, $\hat{h}_{\hat{\mu}\hat{\nu}}\equiv\prescript{(2)}{}{g}_{\hat{\mu}\hat{\nu}}-\eta_{\hat{\mu}\hat{\nu}}\implies\hat{h}_{\mu\nu}\equiv\prescript{(2)}{}{g}_{\mu\nu}-\eta_{\mu\nu}$,
as $\eta_{\mu\nu}$ is invariant under Lorentz transformations.} to the Minkowski metric $\eta_{\mu\nu}$ that gives the metric (\ref{Natario=000020mod})
and the corresponding perturbation for (\ref{Natario=000020mod=000020hat=000020S})
as $\hat{h}_{\mu\nu}$, we now consider the spacetime with metric
\begin{equation}
g_{\mu\nu}=\eta_{\mu\nu}+h_{\mu\nu}+\hat{h}_{\mu\nu}.\label{CTC=000020metric}
\end{equation}
This spacetime contains the two warp drives described by $h_{\mu\nu}$
and $\hat{h}_{\mu\nu}$. To avoid collision of the warp drives, we
make the further assumption that, viewed as functions of $(t,x,y,z)$,
their supports do not overlap:
\[
\supp\bm{\beta}\cap\supp\hat{\bm{\beta}}=\emptyset=\supp(N-1)\cap\supp(\hat{N}-1).
\]
An object travelling inside the first warp drive must be deposited
in the right place in $\hat{S}$ to be picked up by the returning
warp drive. Therefore, the point $(T_{2},\mathbf{r}(T_{2}))$ must
correspond to $(\hat{T}_{0},\hat{\mathbf{r}}(\hat{T}_{0}))$. Using
(\ref{Lor=000020trans}), this means 
\begin{equation}
\begin{gathered}\hat{T}_{0}=\gamma_{u}(T_{2}-uZ(T_{2})),\\
\hat{X}(\hat{T}_{0})=X(T_{2}),\\
\hat{Y}(\hat{T}_{0})=Y(T_{2}),\\
\hat{Z}(\hat{T}_{0})=\gamma_{u}(Z(T_{2})-uT_{2}).
\end{gathered}
\label{midpoint}
\end{equation}
Note that the second warp drive does not have to leave immediately
– we are free to set $\hat{\mathbf{v}}=\mathbf{0}$ for some interval
of time after $\hat{t}=\hat{T}_{0}$. 

We may set similar conditions for the return journey to ensure that
the returning observer is deposited by the second warp drive at exactly
the same spatial coordinates $(x,y,z)=(0,0,0)$ in $S$. Therefore
\[
(T_{finish},\mathbf{0})\leftrightarrow(\hat{T}_{2},\hat{\mathbf{r}}(\hat{T}_{2})),
\]
where $T_{finish}$ is the return time of the observer as measured
in $S$. Thus, we have 
\begin{equation}
\begin{gathered}\hat{T}_{2}=\gamma_{u}T_{finish},\\
\hat{X}(\hat{T}_{2})=0,\\
\hat{Y}(\hat{T}_{2})=0,\\
\hat{Z}(\hat{T}_{2})=-\gamma_{u}uT_{finish}.
\end{gathered}
\label{endpoint}
\end{equation}

\subsection{Requirements for time travel}\label{Sec:=000020CTC=000020req}

Now we define the outgoing and returning average speeds in the $z$
and $-\hat{z}$ directions, as measured in the warp drives' respective
frames: 
\begin{gather}
\bar{v}^{z}\equiv\frac{Z(T_{2})}{T_{2}},\\
\hat{\bar{v}}^{z}\equiv-\frac{\hat{Z}(\hat{T}_{2})-\hat{Z}(\hat{T}_{0})}{\hat{T}_{2}-\hat{T}_{0}}.
\end{gather}
Using (\ref{midpoint}) and (\ref{endpoint}), it is straightforward
to find 
\begin{equation}
T_{finish}=T_{2}\frac{\bar{v}^{z}+\hat{\bar{v}}^{z}-u(1+\bar{v}^{z}\hat{\bar{v}}^{z})}{\hat{\bar{v}}^{z}-u}.\label{T_finish}
\end{equation}
Clearly, if either $\bar{v}^{z}$ or $\hat{\bar{v}}^{z}$ are less
than $1$, we cannot have $T_{finish}<0$. So assuming $\bar{v}^{z},\hat{\bar{v}}^{z}>1$,
we get the \emph{time-travel condition}: 
\begin{equation}
u>\frac{\bar{v}^{z}+\hat{\bar{v}}^{z}}{1+\bar{v}^{z}\hat{\bar{v}}^{z}}.\label{TT=000020condition}
\end{equation}
Let us now assume $T_{finish}<0$. We define the connection between
these two paths at the origin in $S$ 
\begin{equation}
C\equiv\{(t,\mathbf{0}):t\in[T_{finish},0)\},\label{C}
\end{equation}
and note that the entire path given by 
\begin{equation}
\Gamma_{CTC}\equiv\Gamma_{RFT}\cup\hat{\Gamma}_{RFT}\cup C
\end{equation}
is a geodesic, where $\Gamma_{RFT}$ and $\hat{\Gamma}_{RFT}$ are
defined analogously to (\ref{Gamma=000020defn}).

Since the supports of the relevant functions do not overlap, the analysis
from Section \ref{Sec:=000020FT=000020warp=000020drive} still holds.
Therefore, we already know that $\Gamma_{RFT}$ and $\hat{\Gamma}_{RFT}$
are geodesics, and that this path is continuous at the intersections
by (\ref{midpoint}), (\ref{endpoint}), (\ref{C}). In Section \ref{Sec:=000020FT=000020warp=000020drive},
we proved that 
\[
\Gamma_{RFT}=\chi_{1}([0,\tau_{1}))\cup\chi_{2}([\tau_{1},\tau_{2}])
\]
is a geodesic, and a similar proof applies here; the only other thing
we have to check is that the tangent vectors at the start and end
points also match, which indeed turns out to be the case. In $S$,
the observer finishes the outgoing journey with 4-velocity $(\gamma_{u},0,0,u\gamma_{u})$
which, in $\hat{S}$, is $(1,0,0,0)$, agreeing with (\ref{v=000020hat}).
Similarly, the observer finishes the return journey with 4-velocity
$(\gamma_{u},0,0,-u\gamma_{u})$ in $\hat{S}$, which in $S$, is
$(1,0,0,0)$, agreeing with (\ref{C}) and (\ref{v}). Thus $\Gamma_{CTC}$
is a geodesic.

For definiteness, we now also consider the proper time $\tau$ and
its relation to coordinate time $t$. This will give insight into
the points at which the observer is travelling backwards in time,
according to $S$. Within $\Gamma_{RFT}$, we can use the same $t(\tau)$
as we did in Section \ref{Sec:=000020FT=000020warp=000020drive},
that is, 
\begin{equation}
t(\tau)=\begin{cases}
\tau & \tau\in[0,\tau_{1}),\\
\tau_{1}+\epsilon^{-1}(\tau) & \tau\in[\tau_{1},\tau_{2}),
\end{cases}
\end{equation}
where\footnote{$\chi(\tau)$ is again the parametrisation of the outgoing geodesic,
with $(\chi\circ\tau)(t)=(t,\mathbf{r}(t))$.} 
\begin{equation}
\begin{gathered}\epsilon(t)\equiv\int_{T_{1}}^{t}\sqrt{-g(\partial_{t}\chi,\partial_{t}\chi)}\diff t',\qquad t\in[T_{1},T_{2}],\\
\tau_{2}\equiv\tau_{1}+\epsilon(T_{2}).
\end{gathered}
\end{equation}
We can then similarly define $\hat{\epsilon}(\hat{t})$ along with
\begin{equation}
\begin{gathered}\tau_{3}=\tau_{2}+\hat{T}_{1}-\hat{T}_{0},\\
\tau_{4}=\tau_{3}+\hat{\epsilon}(\hat{T}_{2}).
\end{gathered}
\end{equation}
$\tau_{4}$ is the total proper time experienced by the observer in
one round trip along $\Gamma_{CTC}$. This gives us the full description
of the coordinate times $t$ and $\hat{t}$ in terms of the proper
time $\tau$ along $\Gamma_{CTC}$: 
\begin{equation}
\begin{gathered}t(\tau)=\begin{cases}
\tau & \tau\in[0,\tau_{1}),\\
T_{1}+\epsilon^{-1}(\tau) & \tau\in[\tau_{1},\tau_{2}),
\end{cases}\\
\hat{t}(\tau)=\begin{cases}
\hat{T}_{0}+(\tau-\tau_{2}) & \tau\in[\tau_{2},\tau_{3}),\\
\hat{T}_{1}+\hat{\epsilon}^{-1}(\tau) & \tau\in[\tau_{3},\tau_{4}].
\end{cases}
\end{gathered}
\end{equation}
Then, using the reverse Lorentz transformation, the point $(\hat{t},\hat{\mathbf{r}}(\hat{t}))$
in $\hat{S}$ has time coordinate 
\begin{equation}
t=\gamma_{u}(\hat{t}+u\hat{Z}(\hat{t})),
\end{equation}
and we find the full expression for $t(\tau)$: 
\begin{equation}
t(\tau)=\begin{cases}
\tau & \tau\in[0,\tau_{1}),\\
T_{1}+\epsilon^{-1}(\tau) & \tau\in[\tau_{1},\tau_{2}),\\
\gamma_{u}(\hat{T}_{0}+(\tau-\tau_{2})+u\hat{Z}(\hat{T}_{0}+(\tau-\tau_{2}))) & \tau\in[\tau_{2},\tau_{3}),\\
\gamma_{u}(\hat{T}_{1}+\hat{\epsilon}^{-1}(\tau)+u\hat{Z}(\hat{T}_{1}+\hat{\epsilon}^{-1}(\tau)) & \tau\in[\tau_{3},\tau_{4}).
\end{cases}
\end{equation}
As expected, one can show that $\Dot{t}(\tau)>0$ for $\tau\notin(\tau_{2},\tau_{3})$,
and that for $\tau\in(\tau_{2},\tau_{3})$ 
\begin{equation}
\Dot{t}(\tau)<0\iff\partial_{\hat{t}}\hat{Z}(\hat{T}_{0}+(\tau-\tau_{2}))<-\frac{1}{u}<-1,
\end{equation}
so in order to be travelling backwards in time in $S$, the observer
must be moving faster than $\frac{1}{u}$ in $\hat{S}$ in the negative
$\hat{z}$ direction.

\subsection{The curvature tensors}\label{Sec:=000020curvature=000020tensors}

In this section, we give the curvature tensors associated with the
metric (\ref{CTC=000020metric}). Since the metric is defined piecewise
on disjoint regions of $M_{CTC}$, we can safely write 
\begin{equation}
\prescript{(CTC)}{}{\tensor{R}{_{\rho\sigma\mu\nu}}}=\tensor{R}{_{\rho\sigma\mu\nu}}+\tensor{\hat{R}}{_{\rho\sigma\mu\nu}},
\end{equation}
where $\tensor{R}{_{\rho\sigma\mu\nu}}$ and $\tensor{\hat{R}}{_{\rho\sigma\mu\nu}}$
are the Riemann tensors associated to (\ref{Natario=000020mod}) and
(\ref{Natario=000020mod=000020hat=000020S}). 

For this, we shall need the extrinsic curvatures of the spacelike
hypersurfaces defined by constant time $\Sigma_{t}$ and $\hat{\Sigma}_{\hat{t}}$,
and it will be simplest to define them each in their respective frames
as follows: 
\begin{equation}
\begin{gathered}n_{\mu}=(-N,\mathbf{0}),\\
\gamma_{\mu\nu}\equiv\eta_{\mu\nu}+h_{\mu\nu}+n_{\mu}n_{\nu},\\
K_{\mu\nu}\equiv\frac{1}{2}\mathcal{L}_{\mathbf{n}}\gamma_{\mu\nu},
\end{gathered}
\end{equation}
for the outgoing warp drive in $S$ coordinates and 
\begin{equation}
\begin{gathered}\hat{n}_{\hat{\mu}}=(-\hat{N},\mathbf{0}),\\
\hat{\gamma}_{\hat{\mu}\hat{\nu}}\equiv\eta_{\hat{\mu}\hat{\nu}}+\hat{h}_{\hat{\mu}\hat{\nu}}+\hat{n}_{\hat{\mu}}\hat{n}_{\hat{\nu}},\\
\hat{K}_{\hat{\mu}\hat{\nu}}\equiv\frac{1}{2}\mathcal{L}_{\hat{\mathbf{n}}}\hat{\gamma}_{\hat{\mu}\hat{\nu}},
\end{gathered}
\end{equation}
for the returning warp drive in $\hat{S}$ coordinates. 

We shall make use of the Gauss, Codazzi, and Ricci equations\footnote{See, for example, \cite{Bojowald_2010}.},
given here in general: 
\begin{gather}
\tensor{\gamma}{_{\rho}^{\alpha}}\tensor{\gamma}{_{\sigma}^{\beta}}\tensor{\gamma}{_{\mu}^{\gamma}}\tensor{\gamma}{_{\nu}^{\delta}}\prescript{(4)}{}{R}_{\alpha\beta\gamma\delta}=\prescript{(3)}{}{R}_{\rho\sigma\mu\nu}+K_{\rho\mu}K_{\sigma\nu}-K_{\rho\nu}K_{\sigma\mu},\label{Gauss}\\
\tensor{\gamma}{_{\rho}^{\alpha}}n^{\beta}\tensor{\gamma}{_{\mu}^{\gamma}}\tensor{\gamma}{_{\nu}^{\delta}}\prescript{(4)}{}{R}_{\alpha\beta\gamma\delta}=D_{\mu}K_{\rho\nu}-D_{\nu}K_{\rho\mu},\label{Codazzi}\\
\tensor{\gamma}{_{\alpha}^{\rho}}n^{\sigma}\tensor{\gamma}{_{\beta}^{\mu}}n^{\nu}\prescript{(4)}{}{R}_{\rho\sigma\mu\nu}=-\mathcal{L}_{\mathbf{n}}K_{\alpha\beta}+\frac{1}{N}D_{\alpha}D_{\beta}N+K_{\alpha\rho}\tensor{K}{_{\beta}^{\rho}},\label{Ricci}
\end{gather}
where $D$ is the induced covariant derivative on the hypersurface
in question, and $\prescript{(4)}{}{R}_{\rho\sigma\mu\nu},\prescript{(3)}{}{R}_{\rho\sigma\mu\nu}$
are the full and induced Riemann tensors respectively. 

In the following, as in \cite{Santiago_2022}, instead of using the
coordinate basis 
\[
\{\partial_{t},\partial_{x},\partial_{y},\partial_{z}\}
\]
to specify the components of tensors, we shall use the non-coordinate,
orthonormal basis 
\begin{equation}
\{\mathbf{n},\partial_{x},\partial_{y},\partial_{z}\}
\end{equation}
as this will allow us to use Equations (\ref{Gauss})-(\ref{Ricci})
directly. Our notation will be such that a tensor with an index $n$
means that that index has been contracted with $n^{\mu}$, for example
$R_{\mu n}\equiv R_{\mu\rho}n^{\rho}$. 

It turns out that $K_{ij}$ takes a pleasingly simple form: 
\begin{equation}
K_{ij}=\frac{1}{N}\partial_{(i}\beta_{j)},\label{extrinsic}
\end{equation}
and with this in mind, we evaluate $\tensor{R}{_{\rho\sigma\mu\nu}}$.
Since the induced metric is flat, $\gamma_{ij}=\delta_{ij}$, we have
$D_{i}=\partial_{i}$ and $\prescript{(3)}{}{R}_{\rho\sigma\mu\nu}=0$.
Also noting that $\tensor{\gamma}{_{i}^{\rho}}=\tensor{\delta}{_{i}^{\rho}}$,
Equations (\ref{Gauss})-(\ref{Ricci}) give us: 
\begin{equation}
\begin{aligned}R_{ijkl} & =K_{ik}K_{jl}-K_{il}K_{jk},\\
R_{inkl} & =\tensor{\gamma}{_{i}^{\alpha}}n^{\beta}\tensor{\gamma}{_{k}^{\gamma}}\tensor{\gamma}{_{l}^{\delta}}R_{\alpha\beta\gamma\delta}=\partial_{k}K_{il}-\partial_{l}K_{ik},\\
R_{injn} & =\tensor{\gamma}{_{i}^{\rho}}n^{\sigma}\tensor{\gamma}{_{j}^{\mu}}n^{\nu}R_{\rho\sigma\mu\nu}=-\mathcal{L}_{\mathbf{n}}K_{ij}+K_{ik}\tensor{K}{_{j}^{k}}+\frac{1}{N}\partial_{i}\partial_{j}N.
\end{aligned}
\label{Riemann=000020tensor}
\end{equation}
In contracting these expressions, we shall use repeatedly that our
basis is orthonormal, and thus for any tensor $Q_{\mu\nu}$ we may
write 
\[
g^{\mu\nu}Q_{\mu\nu}=-Q_{nn}+\delta^{ij}Q_{ij}.
\]
We now find the Ricci tensor: 
\begin{equation}
\begin{aligned}R_{nn} & =g^{\mu\nu}R_{\mu n\nu n}=-R_{nnnn}+\delta^{ij}R_{injn}\\
 & =-\mathcal{L}_{\mathbf{n}}K-2K_{jk}K^{jk}+K_{jk}K^{jk}+\frac{1}{N}\Delta N\\
 & =-\mathcal{L}_{\mathbf{n}}K-K_{jk}K^{jk}+\frac{1}{N}\Delta N,\\
R_{ni} & =g^{\mu\nu}R_{\mu n\nu i}=-R_{nnni}+\delta^{jk}R_{jnki}\\
 & =\partial_{j}\tensor{K}{_{i}^{j}}-\partial_{i}K,\\
R_{ij} & =g^{\mu\nu}R_{\mu i\nu j}=-R_{ninj}+\delta^{kl}R_{kilj}\\
 & =-\left(-\mathcal{L}_{\mathbf{n}}K_{ij}+K_{ik}\tensor{K}{_{j}^{k}}+\frac{1}{N}\partial_{i}\partial_{j}N\right)+KK_{ij}-K_{ik}\tensor{K}{_{j}^{k}}\\
 & =\mathcal{L}_{\mathbf{n}}K_{ij}+KK_{ij}-2K_{ik}\tensor{K}{_{j}^{k}}-\frac{1}{N}\partial_{i}\partial_{j}N,
\end{aligned}
\end{equation}
where $K\equiv\tensor{K}{_{\rho}^{\rho}}=\tensor{K}{_{i}^{i}}$ and
we have used the useful identity
\begin{equation}
\delta^{ij}\mathcal{L}_{\mathbf{n}}K_{ij}=\mathcal{L}_{\mathbf{n}}K+2K_{ij}K^{ij}\label{extrinsic=000020identity}
\end{equation}
to simplify $R_{nn}$. From this we get the Ricci scalar:
\begin{equation}
\begin{aligned}\label{Ricciscalar}R & =-R_{nn}+\delta^{ij}R_{ij}\\
 & =-(-\mathcal{L}_{\mathbf{n}}K-K_{jk}K^{jk}+\frac{1}{N}\Delta N)+\delta^{ij}(\mathcal{L}_{\mathbf{n}}K_{ij}+KK_{ij}-2K_{ik}\tensor{K}{_{j}^{k}}-\frac{1}{N}\partial_{i}\partial_{j}N)\\
 & =2\mathcal{L}_{\mathbf{n}}K+K^{2}+K_{kl}K^{kl}-\frac{2}{N}\Delta N.
\end{aligned}
\end{equation}
This allows us to evaluate the Einstein tensor $G_{\mu\nu}$, which
is of course related to the energy-momentum tensor by the Einstein
equation 
\[
G_{\mu\nu}=8\pi GT_{\mu\nu}.
\]
We shall give the decomposition of the energy-momentum tensor here:
\begin{equation}
\begin{aligned}\rho\equiv\frac{1}{8\pi G}G_{nn} & =\frac{1}{16\pi G}(K^{2}-K_{ij}K^{ij}),\\
\phi_{i}\equiv\frac{1}{8\pi G}G_{ni} & =\frac{1}{8\pi G}(\partial_{j}\tensor{K}{_{i}^{j}}-\partial_{i}K),\\
T_{ij}\equiv\frac{1}{8\pi G}G_{ij} & =\frac{1}{8\pi G}\big(\mathcal{L}_{\mathbf{n}}K_{ij}+KK_{ij}-2K_{ik}\tensor{K}{_{j}^{k}}-\frac{1}{N}\partial_{i}\partial_{j}N\\
 & \qquad-\frac{1}{2}\delta_{ij}(2\mathcal{L}_{\mathbf{n}}K+K^{2}+K_{kl}K^{kl}-\frac{2}{N}\Delta N)\big).
\end{aligned}
\label{E-M=000020tensor}
\end{equation}

Since Equations (\ref{Riemann=000020tensor}) - (\ref{E-M=000020tensor})
are covariant expressions (if one resets $\partial_{i}\rightarrow D_{i}$),
and $\hat{\gamma}_{\hat{i}\hat{j}}=\delta_{\hat{i}\hat{j}}$ is too,
analogous equations hold for $\hat{R}_{\rho\sigma\mu\nu}$, with $\bm{\beta}\rightarrow\hat{\bm{\beta}}$,
$\mathbf{n}\rightarrow\mathbf{\hat{n}}$, $K_{\mu\nu}\rightarrow\hat{K}_{\mu\nu}$
and $D_{i}\rightarrow\hat{D}_{i}$. 

\subsection{A specific example}\label{Sec:=000020specific=000020example}

\begin{figure}[!h]
\centering
\centering{}\includegraphics[width=1\textwidth]{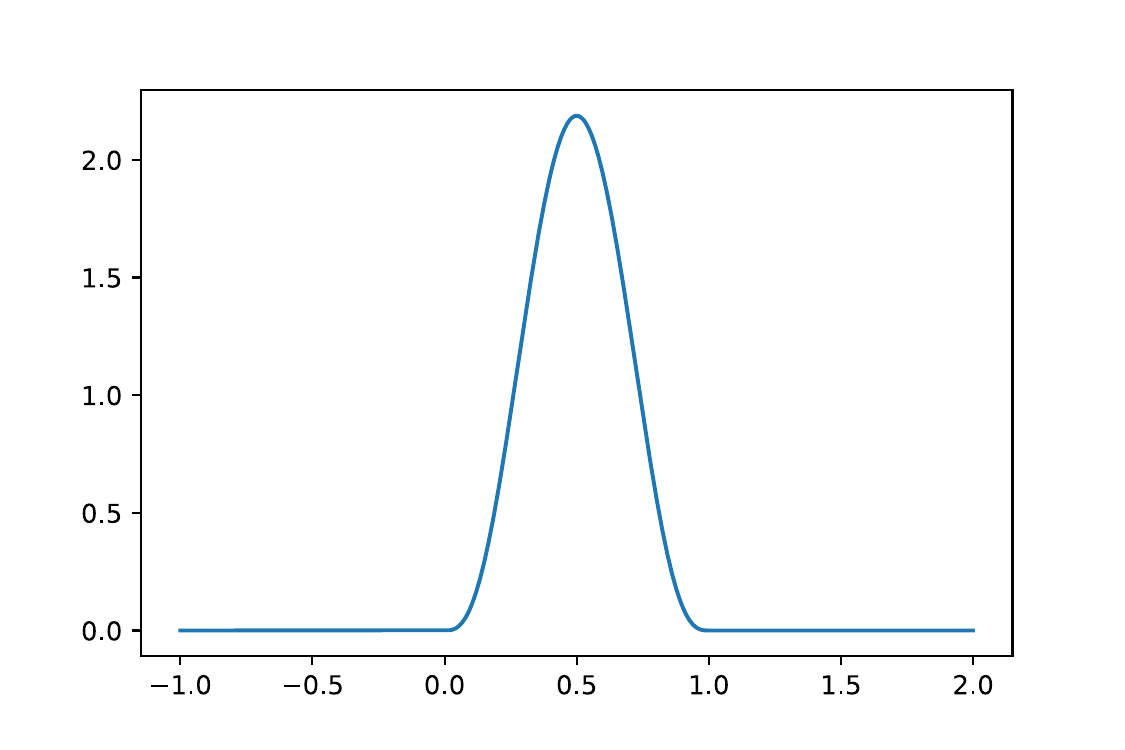}\caption{The bump function $q(x)=140x^{3}(1-x)^{3},\ 0<x<1$.}
\label{Fig:=000020Bump}
\end{figure}

Let us now give a specific choice of the functions that describe our
metric such that the curve $\Gamma_{CTC}$ is indeed a closed timelike
curve, to prove that such a choice is possible and make the result
even more concrete. We first define the bump function 
\begin{equation}
q:\mathbb{R}\rightarrow\mathbb{R},\qquad q(x)=\begin{cases}
140x^{3}(1-x)^{3} & 0\leq x\leq1,\\
0 & \text{otherwise},
\end{cases}
\end{equation}
which can be seen in Figure \ref{Fig:=000020Bump}, and its primitive
\begin{equation}
q^{(-1)}:\mathbb{R}\rightarrow\mathbb{R},\qquad q^{(-1)}(x)=\int_{0}^{x}q(y)\diff y.
\end{equation}
$q^{(-1)}$ is a (non-strictly) monotone-increasing $C^{3}(\mathbb{R})$
function with $q^{(-1)}(x)=0$ for $x\leq0$ and $q^{(-1)}(x)=1$
for $x\geq1$, a `smooth step function'. We shall use $q$ and its
primitives repeatedly for our construction of functions such as $\bm{\beta}$,
$s$, $a$, and $b$. Note that there are many possible choices of
$q$; the following discussion would work just as well with any $q\in C^{2}(\mathbb{R})$
with 
\[
q(x)\in\left[0,1\right],\qquad\supp{q}\subset[0,1],\qquad\textrm{and }\int_{0}^{1}q(x)\diff x=1.
\]

First, we give the functions describing the outgoing warp drive in
$S$. We start with $\mathbf{v}$, the velocity vector field of the
warp drive. We shall take it to be compactly supported, and take
\begin{equation}
r(t,x,y,z)\equiv\big((x-X(t))^{2}+(y-Y(t))^{2}+(z-Z(t))^{2}\big)^{\frac{1}{2}},
\end{equation}
from which we define, similarly to \cite{Alcubierre_1994}, 
\begin{equation}
\begin{gathered}s(r)=1-q^{(-1)}\left(\frac{r-r_{1}}{r_{2}-r_{1}}\right),\\
\mathbf{v}(t,\mathbf{x})=\partial_{t}\mathbf{r}(t)\cdot s(r),
\end{gathered}
\end{equation}
for some $r_{2}>r_{1}$. We see that $s$ satisfies $s(0)=1$ and
$s'(0)=0$. The function $s$ appears in the definition of the lapse
(\ref{lapse}), but here it also serves to define $\mathbf{v}$. $r_{1}$
is the radius of the region of flat space inside the warp bubble as
measured by its passengers, and $r_{2}$ is the radius of the warp
bubble, as measured by external observers in $S$. The direction of
$\mathbf{v}$ does not vary in $(x,y,z)$. 

We wish for the outgoing warp drive to start at rest and finish with
a constant velocity $(0,0,u)$ in $S$. Therefore, the acceleration
along the $z$ axis, $\partial_{t}^{2}Z$, must be zero outside $0<t<T_{2}$,
and it must take both signs at different times to allow for the necessary
acceleration and deceleration. Here, we take it to be comprised of
two bump functions next to each other with opposite signs, as follows:
\begin{equation}
\begin{gathered}\partial_{t}^{2}Z(t)=k_{1}q\left(\frac{t}{t_{1}}\right)-k_{2}q\left(\frac{t-t_{1}}{T_{1}-t_{1}}\right),\\
\partial_{t}Z(t)|_{t=0}=0,\\
Z(0)=0,
\end{gathered}
\end{equation}
for $k_{1},k_{2}>0$ and $0<t_{1}<T_{1}$ where we require 
\begin{equation}
\begin{gathered}\int_{0}^{T_{1}}\partial_{t}^{2}Z(t)\diff t=t_{1}k_{1}-(T_{1}-t_{1})k_{2}=u,\\
\implies t_{1}=\frac{u+T_{1}k_{2}}{k_{1}+k_{2}}.
\end{gathered}
\end{equation}
The warp drive accelerates until $t=t_{1}$, and decelerates from
$t=t_{1}$ to $t=T_{1}$. We get the full solution 
\begin{equation}
Z(t)=k_{1}t_{1}^{2}q^{(-2)}\left(\frac{t}{t_{1}}\right)-k_{2}(T_{1}-t_{1})^{2}q^{(-2)}\left(\frac{t-t_{1}}{T_{1}-t_{1}}\right),
\end{equation}
where
\begin{equation}
\begin{gathered}q^{(-2)}(x)\equiv\int_{0}^{x}q^{(-1)}(y)\diff y.\end{gathered}
\end{equation}
We also define
\[
\kappa\equiv q^{(-2)}(1)=\int_{0}^{1}q^{(-1)}(y)\diff y.
\]
One can calculate that with the above choice of $q$, we have $\kappa=\frac{1}{2}$.
Note that for $x\geq1$, 
\[
q^{(-2)}(x)=\kappa+(x-1).
\]
This gives us 
\begin{equation}
Z(T_{2})=k_{1}t_{1}^{2}\left(\kappa+\frac{T_{2}}{t_{1}}-1\right)-k_{2}(T_{1}-t_{1})^{2}\left(\kappa+\frac{T_{2}-t_{1}}{T_{1}-t_{1}}-1\right).\label{Z(T_2)}
\end{equation}
Next, we choose $a$ such that
\begin{equation}
a(t)=1-q^{(-1)}\left(\frac{t-T_{1}}{t_{2}-T_{1}}\right),
\end{equation}
where $t_{2}=T_{2}-u\gamma_{u}r_{2}$, and we assume $t_{2}>T_{1}$.
The outgoing warp drive vanishes at $t=t_{2}<T_{2}$. This is to avoid
special relativistic issues of simultaneity, where if the outgoing
warp drive were to disappear at $T_{2}$ in $S$ with the returning
warp drive appearing at $\hat{T}_{0}$ in $\hat{S}$, there would
be an overlap (a collision) due to the finite extension of the warp
drives. 

Using (\ref{b=000020from=000020a}), this gives us\footnote{We do not directly prove that the resulting lapse function is positive
everywhere. However, it is clear from (\ref{lapse}) that for a given
$b$, we can choose $r_{2}$ small enough to ensure $N>0$ everywhere.} 
\begin{equation}
b(t)=\frac{u\lambda^{2}}{t_{2}-T_{1}}q\left(\frac{t-T_{1}}{t_{2}-T_{1}}\right),
\end{equation}
where $\lambda$ is defined as before. Note that $b$ is $C^{2}$,
which is important as $b$ forms part of a metric component. The fact
that $q$ is $C^{2}$ implies that the Riemann tensor is at least
$C^{0}$ everywhere.

\begin{figure}[!h]
\centering
\centering{}\includegraphics[width=1\textwidth]{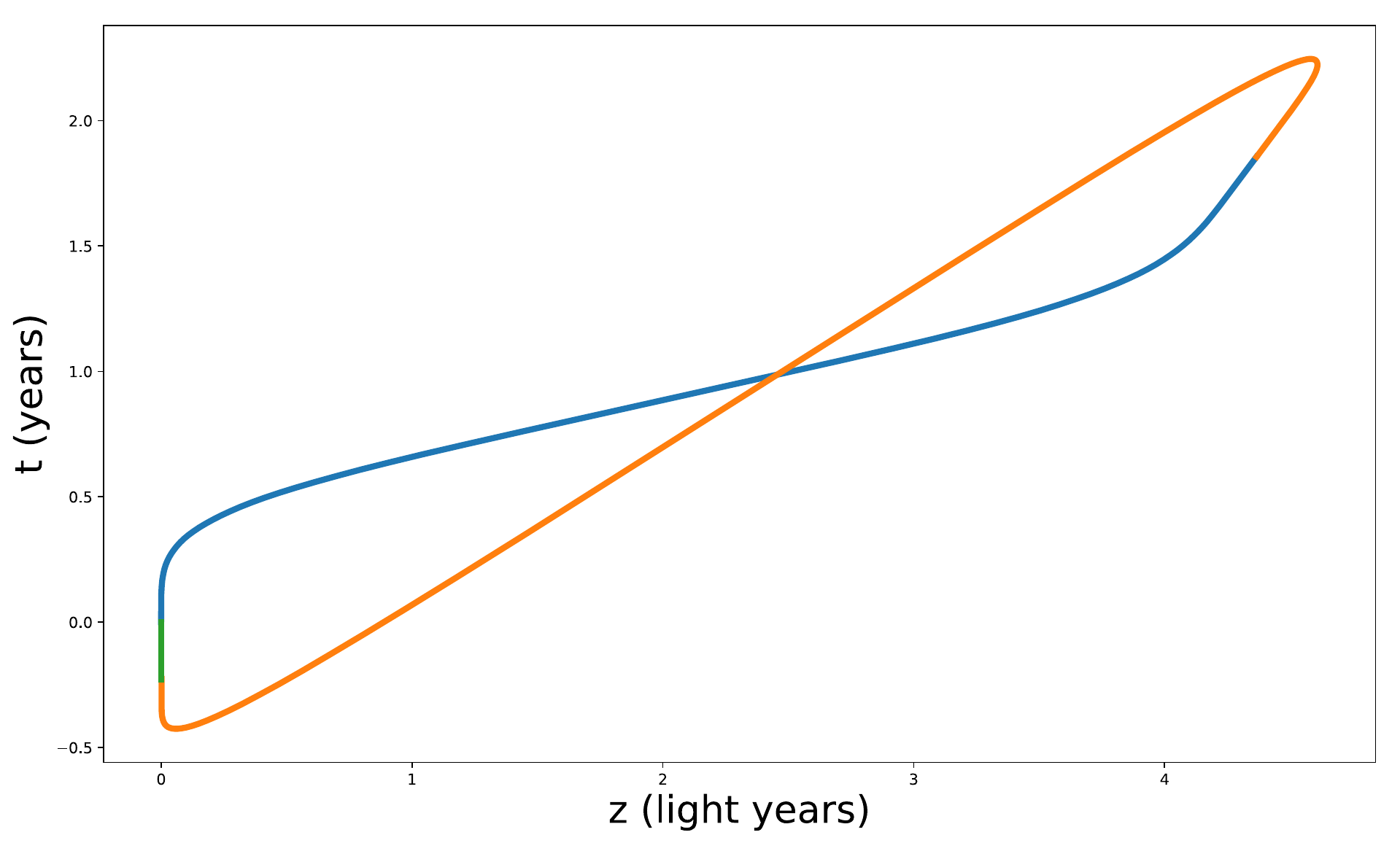}\caption{The path in our example projected onto the $t-z$ plane. The blue
line is the outgoing path, the orange line is the returning path,
and the green line is the section corresponding to $C$. For this
plot we have chosen $u=\frac{3}{4}$, $\alpha=10$, $\omega=0.95$,
and $T=\frac{1}{v}\cdot4.4\ \text{light-years}$. Note that the two
warp drives do not collide in the middle, due to their displacement
in $x$, which is not shown in the plot.}
\label{Fig:=000020t-z=000020diagram}
\end{figure}

Next, we set $X$ and $Y$. It is important that these are not both
simply set to be zero; since the returning warp drive is travelling
backwards in time in $S$ (that is, $\Dot{t}(\tau)<0$), if $X,Y,\hat{X},\hat{Y}$
were simply zero, there would be a collision between the future- and
past-directed warp drives. Therefore, we set 
\begin{equation}
\begin{gathered}X(t)=2r_{2}q^{(-1)}\left(\frac{t}{t_{x}}\right),\\
Y(t)=0,
\end{gathered}
\end{equation}
which allows the outgoing warp drive to `sidestep' the returning
one. We assume $t_{x}\ll T_{2}$, so that the outgoing warp drive
is well out of the way before the returning one gets near. 

Finally, we define the corresponding functions for the returning warp
drive in exactly the same way, but with hats added and various appearances
of $\hat{T}_{0}$. Once again, there are a couple of changes: 
\begin{enumerate}
\item $\partial_{\hat{t}}^{2}\hat{Z}$ is set with an extra minus sign:
\[
\partial_{\hat{t}}^{2}\hat{Z}(\hat{t})=-\hat{k}_{1}q\left(\frac{\hat{t}-\hat{T}_{0}}{\hat{t}_{1}-\hat{T}_{0}}\right)+\hat{k}_{2}q\left(\frac{\hat{t}-\hat{t}_{1}}{\hat{T}_{1}-\hat{t}_{1}}\right),
\]
with $\hat{k}_{1},\hat{k}_{2}>0$. This gives the same condition $(\hat{t}_{1}-\hat{T}_{0})\hat{k}_{1}-(\hat{T}_{1}-\hat{t}_{1})\hat{k}_{2}=u$. 
\item $\hat{X}(\hat{t})=2r_{2}\left(1-q^{(-1)}\left(\frac{\hat{t}-\hat{T}_{0}}{\hat{t}_{\hat{x}}-\hat{T}_{0}}\right)\right)$
(we may leave $r_{1}$ and $r_{2}$ the same for the returning warp
drive). 
\end{enumerate}
To construct an example, we are free to choose these parameters as
we like. We take 
\begin{equation}
\begin{gathered}t_{1}=\frac{1}{2}T_{1},\\
k_{2}=\alpha\frac{u}{T_{1}}\implies k_{1}=(\alpha+2)\frac{u}{T_{1}},
\end{gathered}
\end{equation}
where we leave $\alpha>0$ as a free parameter. Plugging this into
(\ref{Z(T_2)}), and using $\kappa=\frac{1}{2}$, we find 
\begin{equation}
\bar{v}^{z}=\frac{Z(T_{2})}{T_{2}}=u\left(1+\frac{\omega}{4}(\alpha-1)\right),
\end{equation}
where $\omega\equiv\frac{T_{1}}{T_{2}}$. 

We also choose 
\begin{equation}
\begin{gathered}\hat{t}_{1}-\hat{T}_{0}=\frac{1}{2}\left(\hat{T}_{1}-\hat{T}_{0}\right),\\
\hat{k}_{1}=(\alpha+2)\frac{u}{\hat{T}_{1}-\hat{T}_{0}},\\
\hat{k}_{2}=\alpha\frac{u}{\hat{T}_{1}-\hat{T}_{0}},
\end{gathered}
\end{equation}
and set 
\begin{equation}
\frac{\hat{T}_{1}-\hat{T_{0}}}{\hat{T}_{2}-\hat{T}_{0}}=\frac{T_{1}}{T_{2}}=\omega,
\end{equation}
meaning 
\begin{equation}
\bar{v}^{z}=\hat{\bar{v}}^{z}=v.
\end{equation}
Intuitively, these conditions mean that the outgoing path looks the
same in $S$ as the returning path does in $\hat{S}$. This is in
that the profile of the velocity along the $z$ axis has the same
shape, albeit rescaled by a factor $\frac{T_{2}}{\hat{T}_{2}-\hat{T}_{0}}$
to allow the warp drives to travel different distances in their respective
frames. Both warp drives have the same average speed $v$. 

\begin{figure}[!h]
\centering
\centering{}\includegraphics[width=1\textwidth]{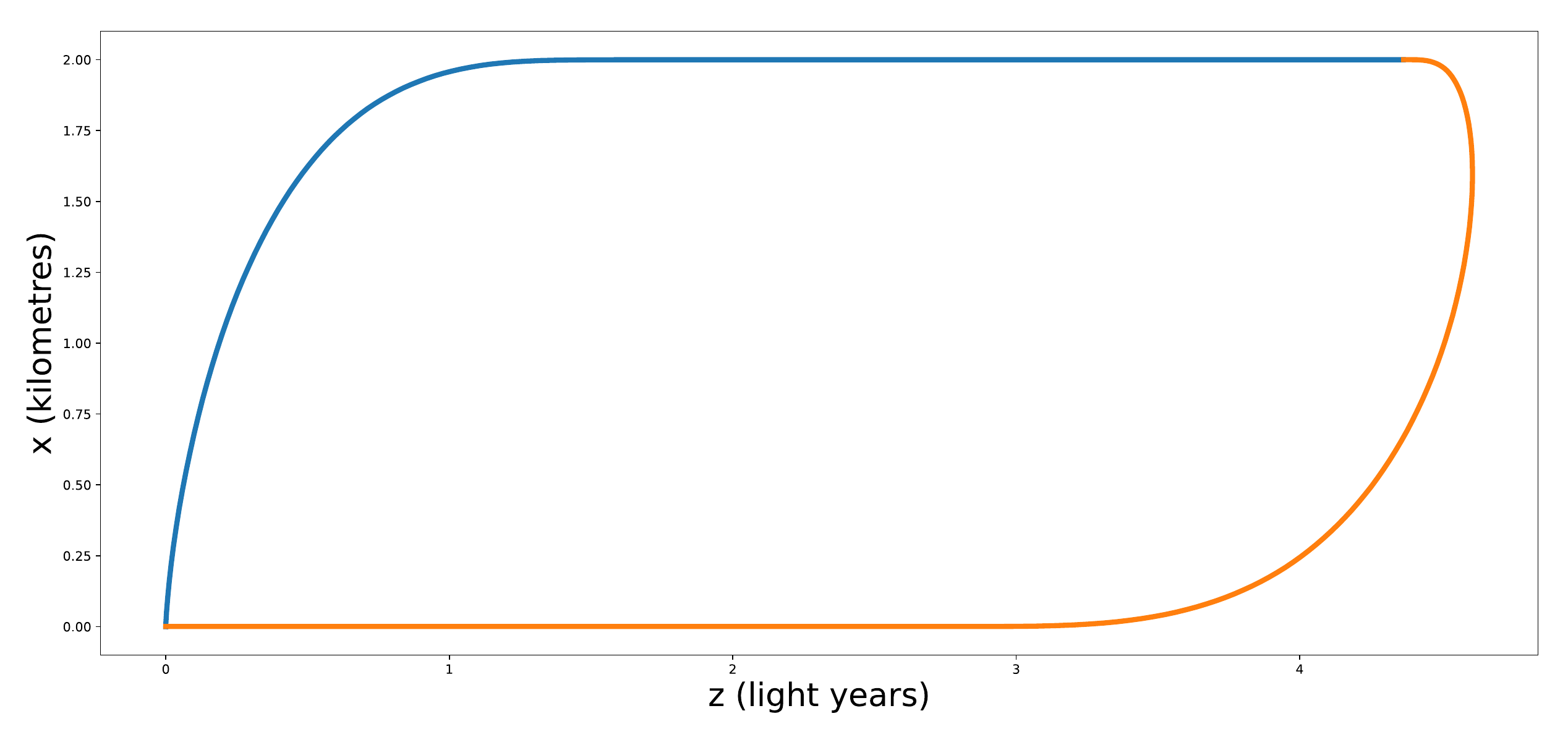}\caption{The path in our example projected onto the $x-z$ plane. The blue
line is the outgoing path, and the orange line is the returning path.
We have made the same choice of variables as in the $t-z$ diagram
in Figure \ref{Fig:=000020t-z=000020diagram}, and in addition took
$t_{x}=\hat{t}_{\hat{x}}-\hat{T}_{0}=0.8\ \text{years}$ and $r_{2}=1\ \text{km}$.
Note that the vertex at $x=z=0$ does not imply the velocity is discontinuous,
since the traveller is stationary in $S$ there.}
\label{Fig:=000020x-z=000020diagram}
\end{figure}

Renaming $T_{2}\to T$ and using (\ref{T_finish}), we get 
\begin{equation}
T_{finish}=T\left(\frac{2v-u(1+v^{2})}{v-u}\right).
\end{equation}
This result can be stated in words as follows: If an observer travels
in a warp drive at an average speed of $v$ in a frame $S$ for a
time $T$, transitions to a frame $\hat{S}$ which is moving at speed
$u$ with respect to $S$, and then returns in another warp drive
moving at speed $-v$ in $\hat{S}$ to the starting point in $S$,
they will arrive at a time $T_{finish}$. If 
\begin{equation}
u>\frac{2v}{1+v^{2}},
\end{equation}
then $T_{finish}<0$, and the observer has travelled back in time.

\section{The weak energy condition with a non-unit lapse}\label{Sec:=000020energy=000020cond}

\subsection{Violations of the weak energy condition}

The study of the energy conditions in relation to warp drive spacetimes
is key, as it gives us a way to determine whether the matter needed
to sustain a given spacetime geometry is `exotic' or not. All warp
drive spacetimes considered in the literature so far violate the weak
energy condition\footnote{Recently, \cite{Lentz_2020}, \cite{Bobrick_2021} and \cite{Fell_2021}
claimed to bypass this limitation. However, these claims were refuted
in \cite{Santiago_2022}, so we will not consider them here.}. Whilst it is still unclear if energy condition violations automatically
mean that the corresponding metric is unphysical, it is usually considered
problematic. This is because known sources of negative energy, for
example the Casimir effect, are believed only to be able to produce
tiny amounts of negative energy, in contrast to the typically vast
negative energy requirements of warp drives.

We shall study the pointwise weak energy condition (WEC) in this section.
In \cite{Santiago_2022}, it was shown that a metric in the Natário
class always entails violation of the WEC, subject to some reasonable
conditions. Here we shall follow a modified version of an argument
in \cite{Santiago_2022} applied to metrics of the form 
\begin{equation}
\diff s^{2}=-N^{2}\diff t^{2}+\delta_{ij}(\diff x^{i}-\beta^{i}\diff t)(\diff x^{j}-\beta^{j}\diff t),\label{gen=000020natario}
\end{equation}
and show that the introduction of a non-unit lapse $N$ does not help
to avoid WEC violation. We also consider the non-trivial case where
$\rho=0$ everywhere, which was not considered in \cite{Santiago_2022}.
Importantly, the arguments in this section hold for an \emph{arbitrary}
positive-definite lapse, not just our particular modification (\ref{lapse})
necessary for creating a closed timelike geodesic.

Our goal will be to study what exactly it is about metrics of the
form (\ref{gen=000020natario}) that gives rise to violation of the
weak energy condition, without consideration for the shape of $N$
and $\bm{\beta}$ (or whether they describe a warp drive in particular).
We do not prove violation of the pointwise null energy condition (NEC)
as was done in \cite{Santiago_2022}, as the arguments therein do
not readily extend to the case with $N\neq1$.

The pointwise weak energy condition (WEC) is the requirement that
\begin{equation}
T_{\mu\nu}t^{\mu}t^{\nu}\geq0\text{ for all timelike vectors }\ensuremath{t^{\mu}\in T_{p}M}.
\end{equation}
The physical justification behind this is that the energy density
at a given point as measured by any timelike observer should be non-negative.
Clearly then, in order to prove WEC violation, only one timelike vector
$t^{\mu}$ such that $T_{\mu\nu}t^{\mu}t^{\nu}<0$ is needed. Here,
we consider the so-called Eulerian energy density $\rho=T_{\mu\nu}n^{\mu}n^{\nu}$.
Using (\ref{extrinsic}) and (\ref{E-M=000020tensor}), we find 
\begin{equation}
\rho=\frac{1}{16\pi G}\frac{1}{N^{2}}((\partial_{i}\beta_{i})^{2}-\partial_{(i}\beta_{j)}\partial_{(i}\beta_{j)}),
\end{equation}
where we are free to write $\bm{\beta}$ with a lowered index since
it is a spatial vector and therefore $\beta_{i}=\delta_{ij}\beta^{j}$.
It is a simple task to rewrite this as 
\begin{equation}
\rho=\frac{1}{16\pi G}\frac{1}{N^{2}}\left(\partial_{i}(\beta_{i}\partial_{j}\beta_{j}-\beta_{j}\partial_{j}\beta_{i})-\partial_{[i}\beta_{j]}\partial_{[i}\beta_{j]}\right),
\end{equation}
and we see that the inside of the brackets is a 3-divergence plus
a negative-semi-definite term. Multiplying by $N^{2}$ and integrating
over a sphere of radius $r$, $B_{t}(r)\subset\Sigma_{t}$, we find
\begin{equation}
\begin{aligned}\int_{B_{t}(r)}N^{2}\rho\diff^{3}x= & \frac{1}{16\pi G}\bigg(\int_{\partial B_{t}(r)}\left(\bm{\beta}(\nabla\cdot\bm{\beta})-(\bm{\beta}\cdot\nabla)\bm{\beta}\right)\cdot\vec{\mathbf{n}}\diff S-\int_{B_{t}(r)}\partial_{[i}\beta_{j]}\partial_{[i}\beta_{j]}\diff^{3}x\bigg),\end{aligned}
\label{int=000020Eulerian}
\end{equation}
where we have used the divergence theorem. The last term on the right-hand
side is clearly non-positive, and the first vanishes as $r\rightarrow\infty$
if we assume that $\bm{\beta}$ decays fast enough. If we take $\bm{\beta}=O(r^{-\frac{1}{2}})$
for large $r$, then 
\begin{equation}
\bm{\beta}(\nabla\cdot\bm{\beta})-(\bm{\beta}\cdot\nabla)\bm{\beta}=O(r^{-2}).
\end{equation}
Therefore, we have 
\begin{equation}
\left(\bm{\beta}(\nabla\cdot\bm{\beta})-(\bm{\beta}\cdot\nabla)\bm{\beta}\right)\cdot\vec{\mathbf{n}}=O(r^{-2}),
\end{equation}
and since $\diff S\propto r^{2}$, the integral vanishes as $r\rightarrow\infty$.
If this is the case, we then find by taking $r\rightarrow\infty$
in (\ref{int=000020Eulerian}) that 
\begin{equation}
\int_{\Sigma_{t}}N^{2}\rho\diff^{3}x\leq0,
\end{equation}
with a strict inequality if $\partial_{[i}\beta_{j]}\neq0$ anywhere,
that is, $\bm{\beta}$ is not curl-free. Then we see that if $\rho>0$
somewhere, $\rho<0$ somewhere else, and the weak energy condition
must be violated somewhere on $\Sigma_{t}$.

\subsection{Discussion and literature review}\label{subsec:Discussion-and-literature}

The argument in the previous section holds if 
\begin{enumerate}
\item The integrand $\bm{\beta}(\nabla\cdot\bm{\beta})-(\bm{\beta}\cdot\nabla)\bm{\beta}$
is $C^{1}$ (continuously differentiable) everywhere such that the
divergence theorem can be applied. This means that $\bm{\beta}$ must
be $C^{2}$, giving a Riemann tensor that is just $C^{0}$ (continuous).\label{req=0000201} 
\item $\bm{\beta}=O(r^{-\frac{1}{2}})$ for large $r$. This can even be
weakened further if, for example, $\bm{\beta}$ is asymptotically
orthogonal to the sphere's normal vector $\vec{\mathbf{n}}$ such
that $\left(\bm{\beta}(\nabla\cdot\bm{\beta})-(\bm{\beta}\cdot\nabla)\bm{\beta}\right)\cdot\vec{\mathbf{n}}=O(r^{-2})$.\label{req=0000202} 
\item $\rho\neq0$ somewhere or $\partial_{[i}\beta_{j]}\neq0$ for some
$i,j$, somewhere. However, if we assume that $\rho=0$ and $\partial_{[i}\beta_{j]}=0$
everywhere, this does not imply that the spacetime is flat or even
that $T_{\mu\nu}=0$. We shall discuss this case further below.\label{req=0000203} 
\end{enumerate}
An interesting feature of the above proof is that it never references
the shape of the spacetime, and certainly does not require that the
spacetime is in any way `warp-drive-like', with a shell of curved
spacetime enclosing a flat interior. Instead, it shows that there
is something fundamental about metrics of this form, subject to these
conditions, that requires the WEC to be violated somewhere. Put another
way, $\Sigma_{t}$ having zero \emph{intrinsic} curvature coupled
with it having non-zero \emph{extrinsic} curvature leads to negative
energy somewhere. 

Although this seems disappointing in terms of the search for a positive-energy
warp drive, this could actually be good news. It tells us that the
negative energy is nothing but a symptom of the choice of metric,
and does \emph{not} have anything to do with \emph{superluminal travel}.
Therefore the WEC violations associated with known warp drives do
not, by themselves, give us any reason to believe that superluminal
travel always requires negative energy. 

An often-cited study \cite{Olum_1998} claims to prove that superluminal
travel always leads to WEC violation. However, this is done using
a definition of superluminal travel based on the idea that a superluminal
path should be in some sense `faster' than all neighbouring paths.
It also relies on the so-called generic condition, which essentially
means the curvature does not vanish entirely along any causal geodesic.\footnote{The generic condition states that along any null geodesic with tangent
vector $\xi^{\mu}$ there is at least one point where 
\begin{equation}
\xi_{[\alpha}R_{\beta]\rho\sigma[\gamma}\xi_{\delta]}\xi^{\rho}\xi^{\sigma}\neq0,
\end{equation}
and along any timelike geodesic with tangent vector $t^{\mu}$ there
is a point where 
\begin{equation}
R_{\alpha\beta\gamma\delta}t^{\alpha}t^{\delta}\neq0.
\end{equation}
}

However, even just the original Alcubierre drive (\ref{Alcubierre})
with $f(x,y,z-\zeta(t))$ chosen such that it is $1$ on an open region
containing $(x,y,z-\zeta(t))=(0,0,0)$ satisfies neither of these
conditions. Indeed, most warp drives considered in the literature
(e.g. \cite{Alcubierre_1994}, \cite{Broeck_1999}, \cite{Nat_rio_2002})
either are or can be easily modified such that a passenger is in a
Riemann-flat environment for the entire journey, so the generic condition
cannot be satisfied.

There will also always be a path just next to the first superluminal
path which is just as fast, so the path is not faster than all its
neighbours. In addition, using the definition of a superluminal path
presented in \cite{Olum_1998} has the consequence that all superluminal
paths are null, which is clearly not the case with even the Alcubierre
drive. The study presented in \cite{Olum_1998} is therefore only
applicable to a limited set of cases which satisfy its particular
definition of superluminal travel.

It it worth noting that if the boundary condition (condition \ref{req=0000202}
above) is not satisfied, one can find an asymptotically-flat $\bm{\beta}$
that goes to zero at spatial infinity, with $\rho\geq0$ everywhere.
It can be shown that, for a scalar $\psi\equiv\psi(t,r)$, where $r=\sqrt{x^{2}+y^{2}+z^{2}}$
and $\bm{\beta}=\nabla\psi$, the Eulerian energy density is given
by 
\begin{equation}
\rho=\frac{1}{8\pi G}\frac{1}{r^{2}}\partial_{r}\left(r(\partial_{r}\psi(t,r))^{2}\right).
\end{equation}
So for example, 
\begin{equation}
\psi(r)=\int_{0}^{r}\frac{s^{\frac{7}{4}}}{(1+s)^{2}}\diff s
\end{equation}
has $\rho\geq0$ everywhere. It is unclear if it is possible to make
$\bm{\beta}$ `warp-drive-like' whilst having $\rho\geq0$ everywhere,
though, and even if it were, one would still need to prove that $T_{\mu\nu}t^{\mu}t^{\nu}>0$
for \emph{all} timelike vectors $t^{\mu}$. This does demonstrate,
however, that negative \emph{Eulerian} energy density is only unavoidable
if the decay condition $\bm{\beta}=O(r^{-\frac{1}{2}})$ is enforced,
and metrics of this form that are only asymptotically flat, for example,
do not \emph{necessarily} violate the WEC. 

In any case, if we restrict ourselves to a sufficiently-smooth and
quickly-decaying vector field $\bm{\beta}$, we can see that the addition
of a non-unit lapse to the Natário class of warp drives does not change
the WEC violation, as was suggested as a possibility in \cite{Santiago_2022}.
This is ignoring the case that $\rho=0=\partial_{[i}\beta_{j]}$ everywhere,
discussed below. 

The integrated Eulerian energy density,
\begin{equation}
E_{tot}=\int_{\Sigma_{t}}\rho\diff^{3}x,
\end{equation}
can however be made arbitrarily small simply by making $N$ large
where $\nabla\bm{\beta}$ is large. If we imagine that $\bm{\beta}$
takes a typical `warp-drive-like' shape with $\nabla\bm{\beta}$
only large inside a thin shell surrounding the passengers, then we
see that $\rho$ is only large inside the shell. Then making $N$
large here can reduce $E_{tot}$ to arbitrarily small values, without
affecting the inside passengers or even the duration of the journey
as measured by them. The trade-off is that this would make the duration
of the journey as experienced by the matter \emph{inside} the shell
larger by a factor of order $N$, as already noted in \cite{Lobo}
(chapter 11, footnote 1).

\subsection{The case $\rho=0=\partial_{[i}\beta_{j]}$}

As mentioned in condition \ref{req=0000203} above, the case where
$\rho=0$ and $\partial_{[i}\beta_{j]}=0$ everywhere does not \emph{necessarily}
violate the WEC. After all, Eulerian observers are just one class
of observers, and even if they see $\rho=0$, this does not mean that
the spacetime is flat or even a vacuum solution. While it is unclear
if this case could ever correspond to anything resembling a warp drive,
it is interesting to see if WEC violations are unavoidable in all
non-trivial metrics of the form (\ref{gen=000020natario}), subject
only to conditions \ref{req=0000201} and \ref{req=0000202}. 

First, we note that 
\begin{equation}
\partial_{[i}\beta_{j]}=\frac{1}{2}\prescript{(3)}{}{\diff}\bm{\beta}\equiv0\iff\exists\,\psi\text{ such that }\prescript{(3)}{}{\diff}\psi=\bm{\beta},
\end{equation}
where $\prescript{(3)}{}{\diff}$ denotes the induced exterior derivative
on $\Sigma_{t}$. Again following arguments in \cite{Santiago_2022},
we know that 
\begin{equation}
\text{WEC}\implies\rho+\bar{p}\geq0,
\end{equation}
where 
\begin{equation}
\bar{p}\equiv\frac{1}{3}\delta^{ij}T_{ij}.
\end{equation}
We can prove this by considering the six vectors in the orthonormal
basis $(\mathbf{n},\partial_{x},\partial_{y},\partial_{z})$:
\begin{equation}
\begin{gathered}(\xi_{1})_{\pm}^{\mu}=(1,\pm a_{1},0,0),\\
(\xi_{2})_{\pm}^{\mu}=(1,0,\pm a_{2},0),\\
(\xi_{3})_{\pm}^{\mu}=(1,0,0,\pm a_{3}),
\end{gathered}
\end{equation}
with $|a_{1}|,|a_{2}|,|a_{3}|<1$ such that $(\xi_{i})_{\pm}$ is
timelike. Since these are all timelike vectors, we have for each $i$:
\begin{equation}
\text{WEC}\implies\begin{gathered}(\xi_{i})_{+}^{\mu}(\xi_{i})_{+}^{\nu}T_{\mu\nu}\geq0,\\
(\xi_{i})_{-}^{\mu}(\xi_{i})_{-}^{\nu}T_{\mu\nu}\geq0.
\end{gathered}
\end{equation}
Expanding these inequalities out and summing them together for each
$i$, the cross terms cancel and we find, since $\rho=T_{\mu\nu}n^{\mu}n^{\nu}$,
\begin{equation}
\text{WEC}\implies2\rho+2a_{i}^{2}T_{ii}\geq0,
\end{equation}
where no summation is implied by the repeated index $i$. Taking the
limit $a_{i}\rightarrow1$, we find 
\begin{equation}
\text{WEC}\implies\rho+T_{ii}\geq0.
\end{equation}
Summing over $i$ and dividing by $3$, we finally see that 
\begin{equation}
\text{WEC}\implies\rho+\frac{1}{3}\delta^{ij}T_{ij}=\rho+\bar{p}\geq0.
\end{equation}
Here we are assuming $\rho=0$, so we have 
\begin{equation}
\text{WEC}\implies\bar{p}\geq0.
\end{equation}
If the 3-momentum tensor $T_{ij}$ is even slightly anisotropic, meaning
its eigenvalues are not all the same, this can be strengthened to
\begin{equation}
\text{WEC }+\text{ anisotropic \ensuremath{T_{ij}}}\implies\bar{p}>0.
\end{equation}
These implications are strictly one-way. 

Let us assume that the WEC holds. Looking at the equation for $\rho$
in (\ref{E-M=000020tensor}), we see 
\begin{equation}
\rho\equiv0\iff K^{2}=K_{ij}K^{ij}.
\end{equation}
Using this, $\beta_{i}=\partial_{i}\psi$, and the equation for $T_{ij}$
in (\ref{E-M=000020tensor}), we see that 
\begin{equation}
\begin{aligned}\bar{p} & =-K^{2}-\mathcal{L}_{\mathbf{n}}K+\frac{1}{N}\Delta N\\
 & =-\frac{1}{N^{2}}(\Delta\psi)^{2}-\frac{1}{N}\left(\partial_{t}\left(\frac{1}{N}\Delta\psi\right)+\partial_{i}\psi\partial_{i}\left(\frac{1}{N}\Delta\psi\right)\right)+\frac{1}{N}\Delta N,
\end{aligned}
\label{bar(p)}
\end{equation}
where we have used $K=\frac{1}{N}\Delta\psi$, and that 
\begin{equation}
\mathcal{L}_{\mathbf{n}}K=n^{\rho}\partial_{\rho}K=\frac{1}{N}(\partial_{t}K+\beta^{i}\partial_{i}K).
\end{equation}
We now multiply by $N$ and integrate over $\Sigma_{t}$. We get 
\begin{equation}
\int_{\Sigma_{t}}N\bar{p}\diff^{3}x=\int_{\Sigma_{t}}\left(-\frac{1}{N}(\Delta\psi)^{2}-\partial_{t}\left(\frac{1}{N}\Delta\psi\right)-\nabla\psi\cdot\nabla\left(\frac{1}{N}\Delta\psi\right)+\Delta N\right)\diff^{3}x.
\end{equation}
The last term vanishes since $\Delta N=\nabla\cdot\nabla N$ is a
divergence, if we make the additional assumption that $N=1+O(r^{-1})$
for large $r$. The condition that $\nabla\psi=\bm{\beta}=O(r^{-\frac{1}{2}})$
is then sufficient to use integration by parts on the penultimate
term. Using this, we find 
\begin{equation}
\begin{aligned}\int_{\Sigma_{t}}N\bar{p}\diff^{3}x & =\int_{\Sigma_{t}}\left(-\frac{1}{N}(\Delta\psi)^{2}-\partial_{t}\left(\frac{1}{N}\Delta\psi\right)-\left(0-\frac{1}{N}(\Delta\psi)^{2}\right)+0\right)\diff^{3}x\\
 & =\int_{\Sigma_{t}}\left(-\partial_{t}\left(\frac{1}{N}\Delta\psi\right)\right)\diff^{3}x=-\int_{\Sigma_{t}}\partial_{t}K\diff^{3}x.
\end{aligned}
\end{equation}
Now, since $N>0$ and the WEC means $\bar{p}\geq0$, we must have
\begin{equation}
\text{WEC}\implies\partial_{t}\int_{\Sigma_{t}}K\diff^{3}x\leq0.\label{irreversible}
\end{equation}
If either $\bar{p}>0$ somewhere or $T_{ij}$ is anisotropic somewhere,
\begin{equation}
\implies\partial_{t}\int_{\Sigma_{t}}K\diff^{3}x<0.
\end{equation}
It is not immediately obvious that this conflicts with anything else
we know about warp drives, but there is one thing we can take from
this. If we start in flat space at $t=0$, we have $K\equiv0$. Therefore,
any warp drive of the form (\ref{Natario=000020mod}) arising from
flat space \emph{must} have, for all $t\geq0$, 
\begin{equation}
\int_{\Sigma_{t}}K\diff^{3}x\leq0,
\end{equation}
and if $\exists\,t>0$ such that $\int_{\Sigma_{t}}K\diff^{3}x<0$,
the spacetime cannot re-flatten without violating the WEC. If $\bar{p}>0$
anywhere or $T_{ij}$ is anisotropic anywhere, this will be the case.
Thus, the only possibility is that $T_{ij}=0$. Then our energy-momentum
tensor becomes 
\begin{equation}
\begin{aligned}\rho & =0,\\
\phi_{i} & =\frac{1}{8\pi G}\left(\partial_{j}\left(\frac{1}{N}(\partial_{i}\partial_{j}\psi)\right)-\partial_{i}\left(\frac{1}{N}\Delta\psi\right)\right),\\
T_{ij} & =0.
\end{aligned}
\end{equation}
Now we can simply take $\phi^{i}=\delta^{ij}\phi_{j}$ and use it
construct the timelike vector 
\[
t^{\mu}=\left(1,-\frac{1}{2}(\phi_{j}\phi^{j})^{-\frac{1}{2}}\phi^{i}\right).
\]
We then calculate 
\begin{equation}
T_{\mu\nu}t^{\mu}t^{\nu}=-(\phi_{j}\phi^{j})^{\frac{1}{2}}\leq0,
\end{equation}
with equality if and only if $\phi^{i}=0$. So the WEC means also
that $\phi_{i}=0$, which altogether shows that $T_{\mu\nu}=0$, so
there is no energy-momentum source. If we have $g_{\mu\nu}=\eta_{\mu\nu}$,
$\partial_{t}g_{\mu\nu}=0$ on $\Sigma_{0}$ (starting from Minkowski
space) and we have $T_{\mu\nu}=0$ for all $t>0$, the unique solution
to Einstein's equations is of course that $g_{\mu\nu}=\eta_{\mu\nu}$
for $t>0$. 

In this paper, we are considering warp drives that appear from and
vanish into flat space, which allows a passenger to exit the warp
bubble. Such warp drives are the most natural ones that could hypothetically
be used for interstellar travel. In this case, there is no chance
this could occur without violating the WEC. The only possible way
for a passenger of such a warp drive to be able to get out is if the
extrinsic curvature scalar $K$ is non-zero only in regions far from
the passenger.

In any case, this certainly forbids simple warp drives with a velocity
parameter $v(t)$ like (\ref{Alcubierre}), since changes in the spacetime
are irreversible\footnote{This does not violate time-reversibility, since $K$ also changes
sign under $t\rightarrow-t$.} by (\ref{irreversible}). This result indicates that although it
may be possible to find a non-trivial metric of the form (\ref{gen=000020natario})
subject to conditions \ref{req=0000201} and \ref{req=0000202} that
does not violate the WEC, it is unlikely that it would be possible
to utilise this to make a useful warp drive metric. 

In conclusion, in this section we have discussed and extended the
results of \cite{Santiago_2022}, and found that the addition of an
arbitrary lapse $N$ does not make a difference with regards to the
WEC violation, although it can be used to reduce the \emph{total}
energy requirements of the warp drive. If we relax condition \ref{req=0000203}
of Section \ref{subsec:Discussion-and-literature}, it is unclear
if WEC violations are unavoidable as a \emph{fundamental} property
of metrics of this form, but we have argued that it is unlikely that
these could actually describe useful positive energy warp drive metrics.

\section{Conclusions}\label{Sec:=000020Summary}

In this paper we provided a concrete, precise, and complete example
of a spacetime geometry which proves that a warp drive can be used
to facilitate not only faster-than-light travel, but also time travel.
While previous authors have already given good arguments for why this
should be possible, the construction we have presented here makes
those arguments precise and gives, for the first time, a complete
double-warp-drive metric that explicitly contains a closed timelike
geodesic.

In order to accomplish this, we generalised the notion of a warp drive
to allow for a non-unit lapse function, a generalisation that has
previously been alluded to in the literature, but never studied in
depth. This was necessary in order to allow the warp drive itself
to transition between two reference frames, without requiring acceleration
from any external forces, which would make the theoretical and mathematical
analysis significantly more complicated.

Using this generalised warp drive, we were able to glue two warp drives
together, and use them to create a closed timelike curve which is
a \emph{geodesic}, a result of the pure spacetime geometry itself.
Our new double-warp-drive spacetime thus joins the existing list of
exotic solutions to Einstein's field equations which contain explicit
closed timelike curves, and it may be used to provide new perspectives
in the study of the nature of time and causality.

We also demonstrated that weak energy condition violations arise in
metrics of the general form 
\begin{equation}
\diff s^{2}=-N^{2}\diff t^{2}+\delta_{ij}(\diff x^{i}-\beta^{i}\diff t)(\diff x^{j}-\beta^{j}\diff t),
\end{equation}
subject to some reasonable conditions, \emph{whether or not} we choose
$N$ and $\bm{\beta}$ such that the spacetime is `warp-drive-like',
that is, contains a warp bubble. Furthermore, these violations occur
whether or not the warp drive is superluminal. This suggests that
the WEC violations arise not from any supposed deep connection between
superluminal travel and negative energy, but rather from the particular
form of the metric. Positive-energy warp drives may still be allowed
within classical general relativity, but only if they could be constructed
using a different type of metric.

We hope that this paper will provide a useful tool for future investigations
of warp drives and their potential use for faster-than-light travel
and/or time travel. In particular, we hope to explore the following
avenues of research in future work:
\begin{itemize}
\item Use the closed timelike geodesics resulting from double-warp-drive
geometry to study various consequences of time travel, such as time
travel paradoxes, in a concrete general-relativistic setting.
\item Check whether our generalised class of non-unit-lapse warp drives
necessarily violates other energy conditions, such as the strong and
null energy conditions.
\item Generalise our warp drive metric even further, by allowing the hypersurfaces
$\Sigma_{t}$ to be intrinsically curved, which may (or may not) allow
us to prevent the WEC violations.
\end{itemize}

\section*{Acknowledgements}

The authors would like to thank Mitacs for funding this research as
part of the Globalink Research Internship program. Ben Snodgrass would
also like to thank Katharina Schramm, Zipora Stober, Alicia Savelli,
and Jared Wogan for stimulating discussions and support, with special
thanks to Jared Wogan for helping with the OGRePy package.

\appendix

\section{In-depth analysis of rest frame transitions}\label{Sec:=000020FT=000020general=000020analysis}

\subsection{The setting}

\begin{figure}[!h]
\centering
\centering{}\includegraphics[width=0.75\textwidth]{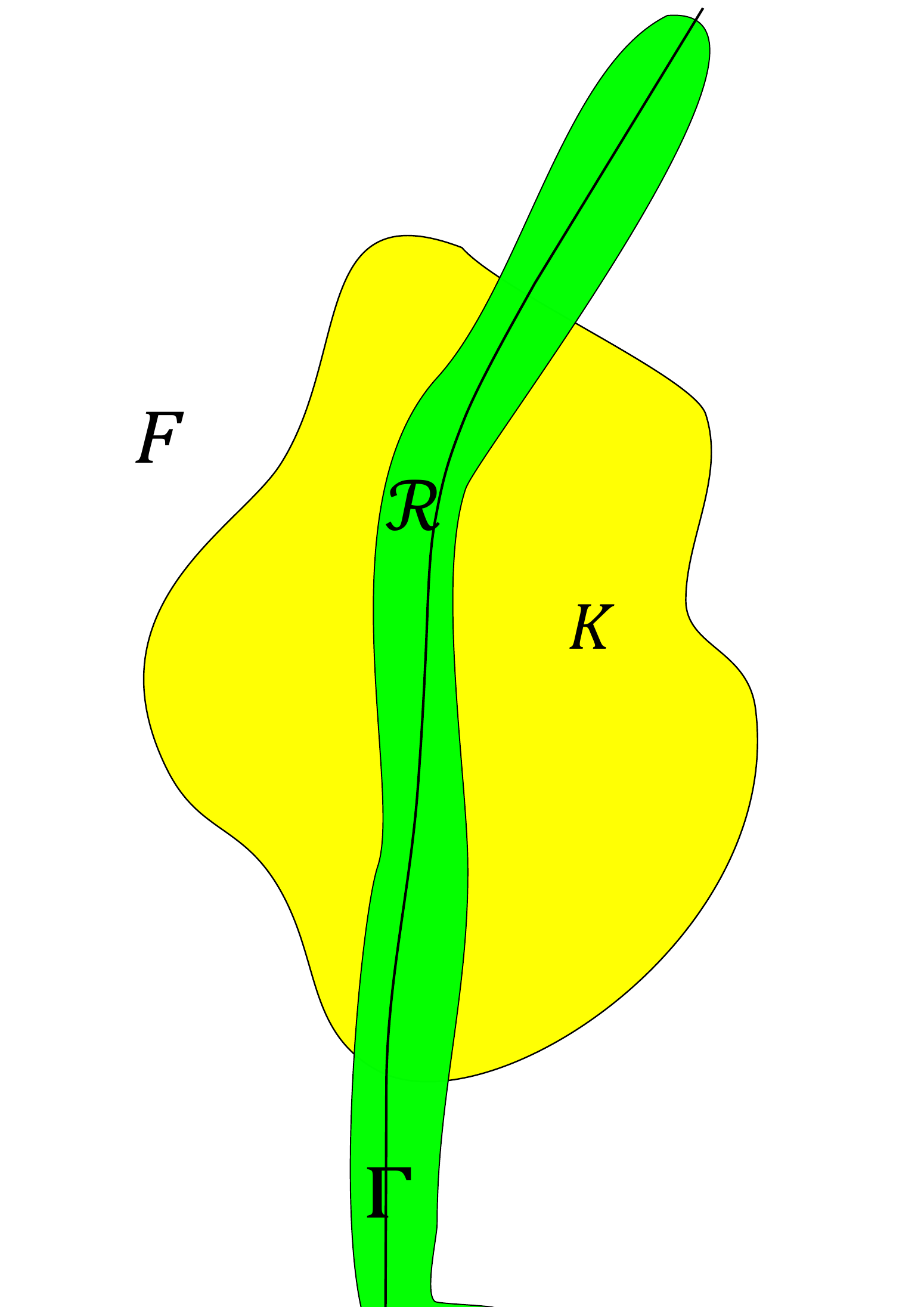}\caption{Schematic diagram showing $\Gamma$, $\mathcal{R}$, $K$ and $F$.}
\label{Fig:=000020WD1}
\end{figure}

In this section we will consider an arbitrary spacetime $M$ subject
to the following conditions, as illustrated in Figure \ref{Fig:=000020WD1}: 
\begin{enumerate}
\item $M$ is diffeomorphic to $\mathbb{R}^{4}$ and geodesically complete. 
\item $M$ is Riemann-flat outside a compact region $K\subset M$, where
$K$ is diffeomorphic to $D^{4}$, the solid 4D ball. We also define
$F\equiv M\setminus K$. 
\item There exists a region $\mathcal{R}\subset M$ diffeomorphic to a 4D
Riemann-flat cylinder of infinite length ($\mathbb{R}\times\textrm{Int}\,D^{3}$),
such that $\mathcal{R}\cap K\neq\phi$ and the intersection $\partial K\cap\mathcal{R}$
is composed of two disjoint 3-surfaces. 
\item $\mathcal{R}$ contains an inextendible timelike geodesic $\Gamma$. 
\end{enumerate}

This setting describes a warp drive that appears from flat space and
subsequently vanishes, returning to flat space. The interior of the
warp drive is $\mathcal{R}$, and the openness of $\mathcal{R}$ ensures
that the flat interior of the warp drive has finite size, and is not
just a single point\footnote{A point-like interior would only be able to transport a point-like
particle without experiencing tidal forces, so this is a reasonable
assumption.}. $\mathcal{R}$ must contain a timelike geodesic in order to allow
a free-falling observer to travel inside the flat interior. This is
a very general conception of a warp drive\footnote{Technically, the CTC spacetimes described in Section \ref{Sec:=000020Making=000020CTC}
would not satisfy these requirements, as the curve $\Gamma_{CTC}$
clearly does not exist inside a region like $\mathcal{R}$, despite
containing two rest frame transitions. This is a mere technicality
though, and to weaken the restrictions on $M$ to allow for this would
increase their complexity significantly, without adding extra physical
insight. Clearly, if one has two compact warp drives subject to the
above conditions, they could be embedded in the same flat spacetime
without issue.}, and in particular makes no assumptions about the existence of a
special time coordinate where, for example, the lapse function is
unity or spacelike hypersurfaces $\Sigma_{t}$ are intrinsically flat. 

The compactness of the curved region $K$ may seem like an unnecessary
and artificial restriction, given that many warp drives studied in
the literature do not obey this constraint (for example, \cite{Alcubierre_1994},
\cite{Nat_rio_2002}, and \cite{Santiago_2022}). However, if we imagine
we are starting in a Riemann-flat universe and creating this warp
drive, information of the warp drive's existence should not be able
to travel arbitrarily fast. The existence of curvature at arbitrary
distances would imply that this information had travelled at arbitrarily
high speeds. 

\subsection{Overview and intuition}

We first give an overview of the arguments and results presented in
this section, before giving their formal proofs. First of all, we
argue that there exists an open sub-region $\tilde{\mathcal{R}}\subset\mathcal{R}$
that is isometric to a `cylinder' of flat spacetime, that is, one
can choose a coordinate system $(\tau,x,y,z)$ in $\tilde{\mathcal{R}}$
such that $g_{\mu\nu}=\eta_{\mu\nu}$, with the boundary of a hypersurface
of constant $\tau$ being a 2-sphere with constant spatial coordinates.
The path $(\tau,\mathbf{0}),-\infty<\tau<\infty$ is $\Gamma$, so
$\Gamma\subset\tilde{\mathcal{R}}$. These would be the local coordinates
used by an observer on the geodesic $\Gamma$. 

We then show that a time coordinate $t$ (which induces a normal vector
field $\mathbf{n}=-\diff t$) can be chosen on $F$ such that $t$
defines a reference frame. An observer following $\Gamma$ is at rest
in this reference frame before entering $K$. Since $F$ extends to
the section of $\Gamma$ beyond $K$, this gives us a well-defined
and natural way to determine whether the observer exits $K$ with
a different 4-velocity from the one they entered it with. 

In Minkowski space, a spacelike hypersurface of zero extrinsic curvature
gives rise to a constant normal vector $\mathbf{n}$, which is the
4-velocity of observers at rest in a frame that has this hypersurface
as the set of points with $t=\text{constant}$. We call a time coordinate
`locally Minkowskian' at a point $p$ if the neighbourhood of $p$
is Riemann-flat and the hypersurface $t=\text{constant}$ is extrinsically
flat in a neighbourhood of $p$.

We next show that there exists an extension of $t$ inside $\tilde{\mathcal{R}}$
such that $t$ is locally Minkowskian everywhere in $\tilde{\mathcal{R}}$
too. Such a choice of $t$ gives a local frame of reference inside
$\tilde{\mathcal{R}}$ and thus has useful intuitive value, as well
as significance in relation to the freedom of choice of the lapse
$N$, as we shall see. It also gives rise to a partial foliation of
$M$, $\Sigma_{t^{*}}\equiv t^{-1}(t^{*})$ for $t^{*}\in\mathbb{R}$. 

After choosing such a $t$, we show that at any point $p\in\tilde{\mathcal{R}}$,
some $\tilde{x},\tilde{y},\tilde{z}$ can be chosen on an open set
containing $p$ where the metric takes the following form: 
\[
\diff s^{2}=-N^{2}\diff t^{2}+\diff\tilde{x}^{2}+\diff\tilde{y}^{2}+\diff\tilde{z}^{2}.
\]
From this, we deduce that since $\tilde{\mathcal{R}}$ is flat, $N$
cannot be of a more general form than 
\[
N(t,\tilde{\mathbf{x}})=p(t)+\mathbf{q}(t)\cdot\tilde{\mathbf{x}},
\]
for some $p$ and $\mathbf{q}$ depending only on time. This was in
fact the original motivation for the ansatz (\ref{lapse}). Lapses
taking this form can be viewed as a uniform `acceleration of the
time coordinate', that is, the integral curve of $n^{\mu}$ would
correspond to an accelerating observer. This means that the non-accelerating
observers in the warp drive accelerate relative to the Eulerian observer
following $n^{\mu}$. Since the vector $n^{\mu}$ is what we measure
the net change in velocity of observers traversing $\tilde{\mathcal{R}}$
against, this corresponds to a rest frame transition. See Figure \ref{Fig:=000020foliation}
for a visual depiction of this.

\subsection{Formalism and proofs}

\subsubsection{Construction of $\tilde{\mathcal{R}}$}

First, we prove the existence of $\tilde{\mathcal{R}}$ as described
above. Take a point $p_{0}\in\Gamma$ in the causal past of $\Gamma\cap\partial K$
(i.e. before the observer enters the warp drive) and its timelike
tangent vector along $\Gamma$, $\mathbf{n}_{0}\in T_{p_{0}}M$. Since
$\mathcal{R}$ is simply connected and Riemann-flat, a coordinate
system $(\tau,x,y,z):\mathcal{R}\rightarrow\mathbb{R}$ can be constructed
such that associated tangent vector to the coordinate $\tau$ has
the same components as $\mathbf{n}_{0}$, and that the metric takes
the Minkowskian form\footnote{$\tau$ has been named as such since it corresponds to the proper
time of a passenger following $\Gamma$.}. $\Gamma$ is the path $(\tau,\mathbf{0}),\tau\in\mathbb{R}$. Now
define $\tilde{r}>0$ by 
\begin{equation}
\tilde{r}^{2}\equiv\inf{\{x^{2}(q)+y^{2}(q)+z^{2}(q):q\in\partial\mathcal{R}\}}.
\end{equation}
By openness of $\mathcal{R}$ and compactness of $K$, we must have\footnote{Technically, $\mathcal{R}$ could become arbitrarily thin as $\tau\rightarrow+\infty$
or $\tau\rightarrow-\infty$, but since this is safely far away from
$K$, one could simply choose $\mathcal{R}'\supset\mathcal{R}$ such
that this is not the case.} $\tilde{r}>0$. Then we simply define 
\begin{equation}
\tilde{\mathcal{R}}\equiv\{q\in\mathcal{R}:x^{2}(q)+y^{2}(q)+z^{2}(q)<\tilde{r}^{2}\},
\end{equation}
and our construction is complete. $\tilde{\mathcal{R}}$ is a subset
of $\mathcal{R}$, is clearly open as it is the preimage of an open
set under a continuous mapping, and in these coordinates, it is manifest
that $\tilde{\mathcal{R}}$ is isometric (not just homeomorphic like
$\mathcal{R}$) to an infinitely-long, Riemann-flat 3+1 cylinder.

\subsubsection{Construction of the time coordinate and reference frame outside $K$}

\begin{figure}[!h]
\centering
\centering{}\includegraphics[width=0.75\textwidth]{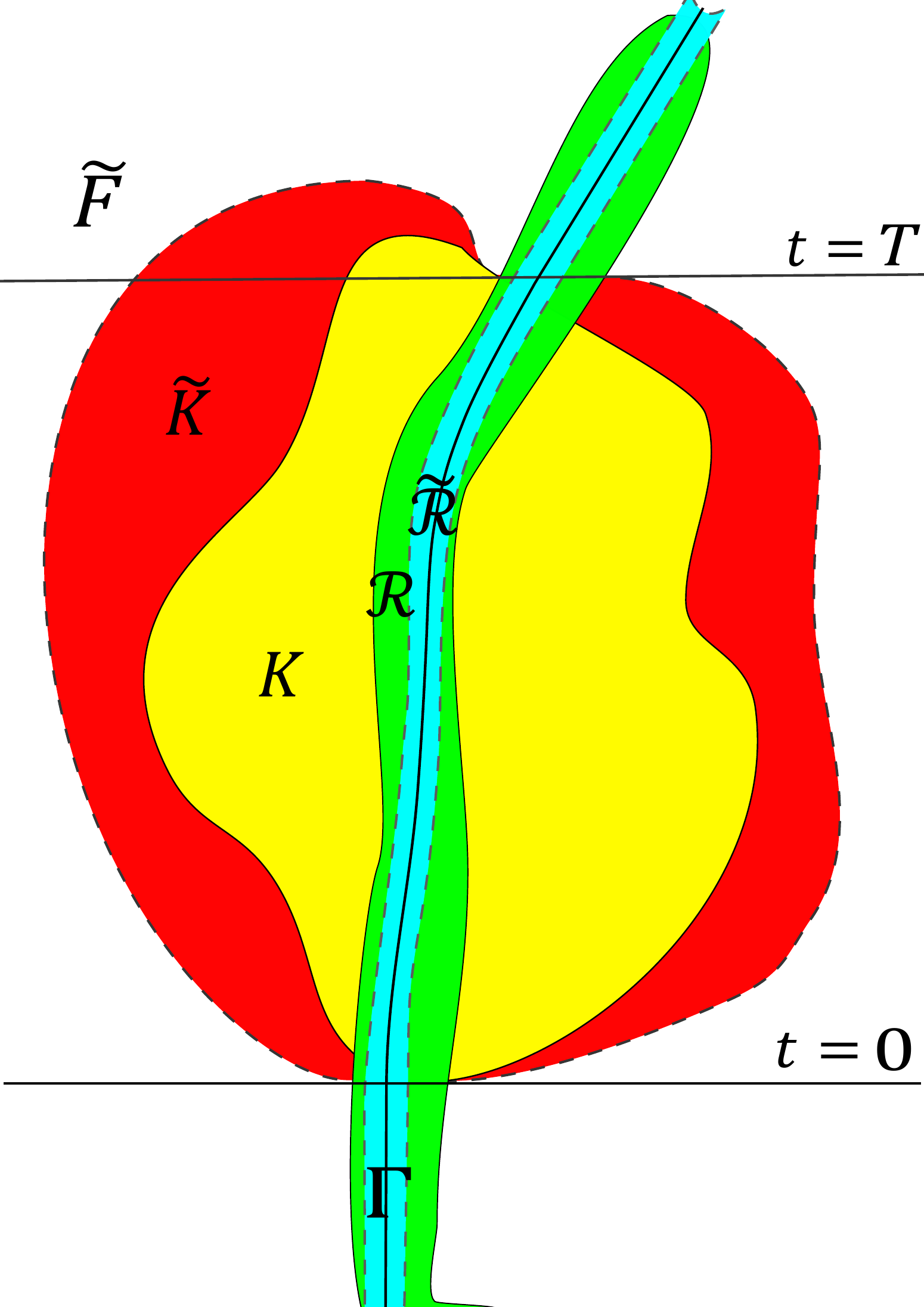}\caption{Schematic diagram showing $\Gamma$, $\mathcal{R}$, $\tilde{\mathcal{R}}$,
$K$, $\tilde{K}$ and $\tilde{F}$. $\tilde{K}\supset K$ is such
that the boundary $\partial\tilde{K}\cap\tilde{\mathcal{R}}$ is composed
of two sections, one at $t=0$ and the other at $t=T$, the \textquoteleft entry\textquoteright{}
and \textquoteleft exit\textquoteright{} of $\tilde{K}$.}
\label{Fig:=000020WD2}
\end{figure}

For a schematic diagram displaying the following decomposition of
$M$, see Figure \ref{Fig:=000020WD2}. A reference frame, in Minkowski
space, can be determined from the choice of only one timelike vector
at some point, and extending it by parallel transport to make a vector
field $\mathbf{n}$ on the whole of Minkowski space. Since Minkowski
space is flat and simply connected, this extension is unique. There
will also exist a timelike coordinate $t$ that one can derive from
this, satisfying $-\diff t=\mathbf{n}$, with the minus sign a convention
ensuring that $\mathbf{n}$ is future-pointing, that is, $n^{0}>0$. 

To create a global reference frame on the flat region $F$, we can
similarly take the tangent vector of $\Gamma$, $\mathbf{n}_{0}$,
at $p_{0}$, and define a vector at another point $q\in F$ by parallel-transporting
the vector at $p$ along a path contained in $F$. Since $F$ is Riemann-flat
and simply connected, the resulting vector at $q$ is independent
of the path chosen. Using this method, we can define a unique vector
field 
\[
\mathbf{n}:p\in F\rightarrow T_{p}F.
\]
This gives us a global reference frame on $F$, set to be at rest
with respect to the observer following $\Gamma$ \emph{before} entering
$K$, which allows us to quantify the change in the observer's 4-velocity
after passing through $K$, thus determining whether a rest frame
transition has occurred. Again, we can take a coordinate $t:F\rightarrow\mathbb{R}$
such that $-\diff t=\mathbf{n}$, choosing $t$ such that 
\begin{equation}
0=\inf{\{t(\tilde{\mathcal{R}}\cap K)\}},
\end{equation}
and we define 
\begin{equation}
T\equiv\sup{\{t(\tilde{\mathcal{R}}\cap K)\}}.
\end{equation}
Finally, we choose some $\tilde{K}\supset K$ such that the intersection
of $\tilde{\mathcal{R}}$ and $\partial\tilde{K}$ has $t(\tilde{\mathcal{R}}\cap\partial\tilde{K})=\{0,T\}$,
that is, we extend $K$ to $\tilde{K}$ so that the boundary of the
section of $\tilde{\mathcal{R}}$ contained in $\tilde{K}$ has $t=0$
at one end and $t=T$ at the other. We also de-specify $t$ on $\tilde{K}\setminus K$
so that we are free to choose $t$ as we like inside $\tilde{\mathcal{R}}\cap\tilde{K}$.
$\tilde{\mathcal{R}}$ and $\tilde{K}$ will be the main regions we
shall study from now on. 

It is crucial that we do not extend $\mathbf{n}$ along a path inside
$\mathcal{R}$, as the point of this analysis is to see how the 4-velocity
of an observer following $\Gamma$ changes relative to the \emph{background}
reference frame defined by $\mathbf{n}$ in $F$. If we extended it
along a path \emph{inside} $\mathcal{R}$, we would ensure that the
observer following $\Gamma$ remains at rest with respect to the local
reference frame defined by $\mathbf{n}$, and the possible discontinuity
in the normal vector field from extending $\mathbf{n}$ to $\mathcal{R}\cup F$
would instead occur in $F$.

\subsubsection{Construction of the time coordinate inside $\tilde{\mathcal{R}}$}

In $\tilde{F}\equiv M\setminus\tilde{K}$, the time coordinate is
already specified. How may we choose it inside $\tilde{K}$, and in
particular $\tilde{\mathcal{R}}\cap\tilde{K}$? One may imagine that
since $\tilde{\mathcal{R}}$ is flat, one can analytically extend
$t$ inside $\tilde{\mathcal{R}}\cap\tilde{K}$, to create a continuous
time coordinate with a constant associated normal vector inside $\tilde{\mathcal{R}}$.
However, this is not in general possible, a consequence of the flat
region $\tilde{F}\cup\tilde{\mathcal{R}}$ not being simply connected.
There is no guarantee that the parallel transport of $\mathbf{n}$
defined at some point with $t<0$ along a path traversing $\tilde{\mathcal{R}}$
will match $\mathbf{n}$ on the other side of $K$. 

So what freedom do we have to choose $t$ inside $\tilde{\mathcal{R}}$?
What is the most natural choice of $t$? If it were \emph{always}
possible to choose a time coordinate $t$ such that within $\tilde{\mathcal{R}}$
we have $N=N(t)$, this would imply that rest frame transitions are
impossible\footnote{At least for warp drives with a finite region of flat space inside
them, that is, warp drives with a non-point-like interior.}, contrary to the example in Section \ref{Sec:=000020FT=000020warp=000020drive}.
Therefore, we must search for something weaker. The next-best option
is to have $\tilde{R}_{t}\equiv\tilde{\mathcal{R}}\cap\Sigma_{t}$
extrinsically flat (a locally-Minkowskian time coordinate), which
in turn implies that it is also intrinsically flat by the Gauss equation.
This is given in general as follows: 
\begin{equation}
\tensor{\gamma}{_{\rho}^{\alpha}}\tensor{\gamma}{_{\sigma}^{\beta}}\tensor{\gamma}{_{\mu}^{\gamma}}\tensor{\gamma}{_{\nu}^{\delta}}\prescript{(4)}{}{R}_{\alpha\beta\gamma\delta}=\prescript{(3)}{}{R}_{\rho\sigma\mu\nu}+K_{\rho\mu}K_{\sigma\nu}-K_{\rho\nu}K_{\sigma\mu},\label{Gauss=000020eqn}
\end{equation}
where $\gamma_{\mu\nu}=g_{\mu\nu}+n_{\mu}n_{\nu}$. Since the ambient
space $\tilde{\mathcal{R}}$ is flat and the extrinsic curvature $K_{\mu\nu}$
is assumed to vanish, we are left with 
\[
\prescript{(3)}{}{R}_{\rho\sigma\mu\nu}=0.
\]

We will now show that such a choice of $t$ always exists by explicit
construction. Once such a $t$ has been chosen, we will be able to
see exactly what the restrictions on the associated lapse $N$ are. 

Take $p_{0}$ again and, using the same method with parallel transport,
construct a vector field on $\tilde{\mathcal{R}}$ from the normal
vector $\mathbf{n}$ at this point: 
\[
\mathbf{n}_{1}:p\in\tilde{\mathcal{R}}\rightarrow T_{p}\tilde{\mathcal{R}}.
\]
In exactly the same way, choosing some point in $\Gamma$ after it
has traversed $\tilde{K}$, we can construct a second vector field:
\[
\mathbf{n}_{2}:p\in\tilde{\mathcal{R}}\rightarrow T_{p}\tilde{\mathcal{R}}.
\]
These vector fields are the extension of the global reference frame
on $\tilde{F}$ inside $\tilde{\mathcal{R}}$, with $\mathbf{n}_{1}$
from the past of $\tilde{K}$ and $\mathbf{n}_{2}$ from the future.
As explained above, there is no reason that they must agree inside
$\tilde{\mathcal{R}}$. 

Now we reuse the Minkowskian coordinate system $(\tau,x,y,z)$ inside
$\tilde{\mathcal{R}}$, choosing $\tau=0$ to coincide with $t=0$.
In the coordinate basis we have $n_{1}^{\mu}=(1,\mathbf{0})$, and
we now define $\bm{\sigma}$ such that 
\begin{equation}
\begin{gathered}n_{2}^{\mu}=(\gamma,\gamma\bm{\sigma}),\\
\gamma\equiv(1-\sigma^{2})^{-\frac{1}{2}},\qquad\sigma\equiv|\bm{\sigma}|<1.
\end{gathered}
\end{equation}
$\bm{\sigma}$ is a constant vector, so $\diff\bm{\sigma}=0$, where
$\diff$ denotes the exterior derivative. This implies $\exists\thinspace\phi:\tilde{\mathcal{R}}\rightarrow\mathbb{R}$
such that $\bm{\sigma}=\diff\phi$. Looking at the expressions for
$\mathbf{n}_{1},\mathbf{n}_{2}$ and $\bm{\sigma}$, we see 
\begin{equation}
\mathbf{n}_{2}=\gamma(\mathbf{n}_{1}+\bm{\sigma})\iff\mathbf{n}_{2}=\gamma(-\diff\tau+\diff\phi),\label{normal=000020relation}
\end{equation}
a coordinate-independent equation. Now define 
\[
\tau_{0}\equiv\inf{\{\tau(p):p\in\tilde{\mathcal{R}}\cap t^{-1}(T)}\},
\]
recalling that 
\[
\tilde{\mathcal{R}}\cap t^{-1}(T)\subset\partial\tilde{K},
\]
that is, $\tilde{\mathcal{R}}\cap t^{-1}(T)$ is the `exit' of the
compact region $\tilde{K}$ inside $\tilde{\mathcal{R}}$. $\phi$
is only specified up to an arbitrary constant, so we take it that
$\phi=0$ somewhere in the boundary of $\tilde{R}_{T}$, with $\phi\geq0$
inside $\tilde{R}_{T}$. 

Consider the section of the hypersurface $t=T$ inside $\tilde{\mathcal{R}}$,
the `exit' of $\tilde{K}$. This has a constant normal 
\[
(n_{2})_{\mu}=(-\gamma,\gamma\bm{\sigma}),
\]
so using usual Euclidean geometry, it has the planar equation 
\begin{equation}
-\tau+\bm{\sigma}\cdot\mathbf{x}=\text{constant}.
\end{equation}
Thus, we see that the infimum of $\tau$ on this surface, $\tau_{0}$,
will be reached at the (limiting) point where $\bm{\sigma}\cdot\mathbf{x}$
is minimised. Since $\diff\phi=\bm{\sigma}$, we can also see that
\begin{equation}
\phi=\phi_{0}+\bm{\sigma}\cdot\mathbf{x},
\end{equation}
for some $\phi_{0}\in\mathbb{R}$. Therefore we see that minimising
$\tau$ within the surface $t=T$ is equivalent to minimising $\phi$.
Since we set $\phi$ such that its minimum on $\tilde{R}_{T}$ is
zero, for any point where $\tau=\tau_{0}$ and $t=T$, $\phi=0$.
Also, since hypersurfaces of constant $\tau$ have a fixed boundary
in these coordinates (independent of $\tau$) and $\phi$ does not
depend on $\tau$ either, this means $\phi\geq0$ everywhere inside
$\tilde{\mathcal{R}}$. 

\begin{figure}[!h]
\centering
\centering{}\includegraphics[width=0.5\textwidth]{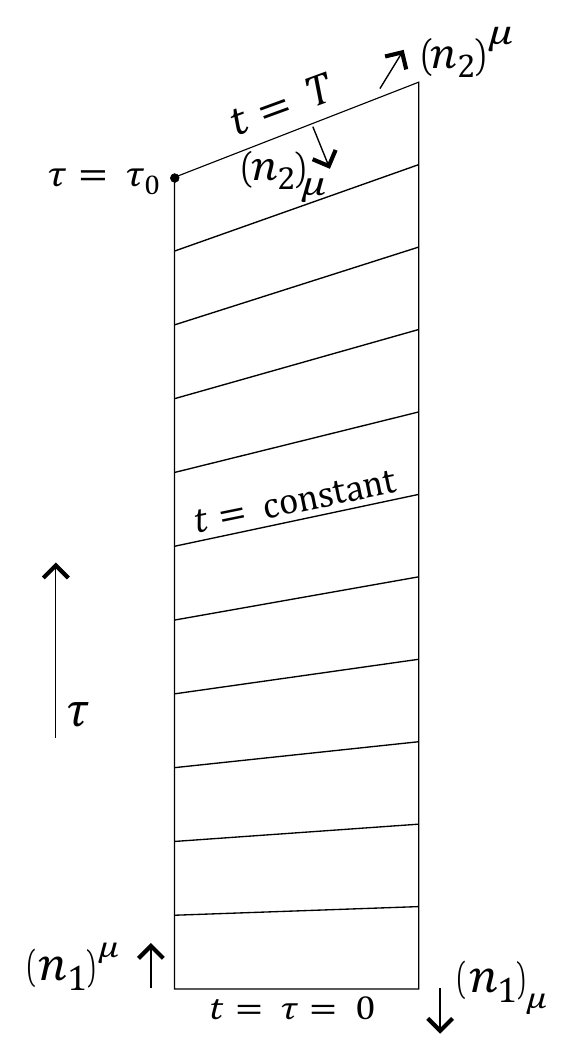}\caption{A choice of $t$ that is locally Minkowskian, in the region $\tilde{\mathcal{R}}\cap\tilde{K}$.
The vector fields $\mathbf{n}_{1}$ and $\mathbf{n}_{2}$ have constant
components throughout $\tilde{\mathcal{R}}$.}
\label{Fig:=000020foliation}
\end{figure}

Now it is time to choose the time coordinate $t$ inside $\tilde{\mathcal{R}}$.
There are many possible choices that give a valid time coordinate,
continuous at $t=0$ and $t=T$ with $\tilde{R}_{t}$ being extrinsically
flat, so here we give just one, to prove existence: 
\begin{equation}
t=\frac{T}{\tan{\sigma}}\tan{\left(\frac{\sigma\tau}{\tau_{0}+\phi}\right)},\label{t=000020choice}
\end{equation}
or if $\sigma=0$, we simply take 
\begin{equation}
t=T\left(\frac{\tau}{\tau_{0}}\right),\label{t=000020choice=0000200=000020sigma}
\end{equation}
as in that case $\phi\equiv0$. See Figure \ref{Fig:=000020foliation}
for a diagram showing this choice of $t$. This satisfies the necessary
conditions: 
\begin{enumerate}
\item $\tau=0\implies t=0$ and ($\tau=\tau_{0},\phi=0)\implies t=T$. 
\item The hypersurfaces $\tilde{R}_{t}$ are extrinsically flat. 
\item $t=T\implies-\diff t$ is parallel to $\mathbf{n}_{2}$, and thus
the \emph{normalised} normal vector field associated to this choice
of $t$ is continuous, and $t$ is a legitimate time coordinate. 
\end{enumerate}
The first of these points follows immediately from (\ref{t=000020choice})
and (\ref{t=000020choice=0000200=000020sigma}). The second can be
seen by observing that a hypersurface in $\tilde{\mathcal{R}}$ with
$t=\text{constant}$ is described by 
\begin{equation}
\tau=k(\tau_{0}+\phi),
\end{equation}
for some $k>0$. Since $\phi$ takes the form $\phi=\phi_{0}+\bm{\sigma}\cdot\mathbf{x}$,
$\tilde{R}_{t}$ is isometric to a section of a 3-dimensional plane
in Minkowski space, and clearly has no extrinsic curvature. The last
point follows from calculating $-\diff t$, and finding 
\[
-\diff t\propto-(\tau_{0}+\phi)\diff\tau+\tau\diff\phi,
\]
where the constant of proportionality is strictly positive. Using
this and (\ref{normal=000020relation}), the condition for $-\diff t\propto\mathbf{n}_{2}$
is that $\tau_{0}+\phi=\tau$. But this is precisely the condition
that $t=T$ in (\ref{t=000020choice}) and (\ref{t=000020choice=0000200=000020sigma}).
Therefore, our time coordinate $t$ is continuous with a continuous,
normalised, everywhere-timelike normal vector field $\mathbf{n}=-N\diff t$,
where $N:\tilde{F}\cup\tilde{\mathcal{R}}\rightarrow(0,\infty)$ is
the associated lapse, and the construction of our time coordinate
is complete.

\subsubsection{Construction of local coordinate system in $\tilde{\mathcal{R}}$}

Now that we have vanishing extrinsic curvature of $\tilde{R}_{t}$,
it is time to see what this means. Choose a $(x',y',z')$ on $\tilde{\mathcal{R}}$
such that in the ADM formalism, its metric can be written as 
\begin{equation}
\diff s^{2}=-N^{2}\diff t^{2}+\delta_{ij}(\diff x'{}^{i}-\beta'{}^{i}\diff t)(\diff x'{}^{j}-\beta'{}^{j}\diff t),
\end{equation}
for some shift vector $\bm{\beta}'$. This must be possible since
the spatial slices are intrinsically flat, so we can take the induced
metric to be $\delta_{ij}$. The extrinsic curvature tensor of a hypersurface
$t=\text{constant}$ in this metric is
\begin{equation}
K_{ij}=\frac{1}{N}\partial_{(i}\beta'_{j)}.\label{extrinsic0}
\end{equation}
Therefore, the vanishing of the extrinsic curvature implies that $\bm{\beta}'$
is a Killing vector of the hypersurfaces. Now we apply the coordinate
transformation 
\begin{gather*}
t\rightarrow t,\\
x'\rightarrow\tilde{x}(t,x',y',z'),\\
y'\rightarrow\tilde{y}(t,x',y',z'),\\
z'\rightarrow\tilde{z}(t,x',y',z'),
\end{gather*}
to some small open region inside $\tilde{\mathcal{R}}$, setting 
\begin{equation}
\begin{gathered}(\partial_{t}\tilde{x},\partial_{t}\tilde{y},\partial_{t}\tilde{z})=\bm{\beta}',\\
(\tilde{x}(t_{0},x',y',z'),\tilde{y}(t_{0},x',y',z'),\tilde{z}(t_{0},x',y',z'))=(x',y',z'),
\end{gathered}
\end{equation}
for some $0<t_{0}<T$. The region must be small in order to ensure
that the integral curves of $\bm{\beta}'$ do not go so far as to
leave $\tilde{\mathcal{R}}$. However, since this argument can be
applied to any part of $\mathcal{\tilde{R}}$, this is not problematic.
One can calculate that this means the new metric $g_{\mu\nu}$ has
$g_{0\mu}=(-N^{2},0,0,0)$. Without calculation, we also know that
the spatial metric must be preserved, as we are shifting points along
a Killing vector field of the hypersurfaces, under which the spatial
metric is invariant.

Noting that since the time coordinate $t$ did not change, the lapse
$N$ did not change either, we see that the metric now becomes 
\[
\diff s^{2}=-N^{2}\diff t^{2}+\diff\tilde{x}^{2}+\diff\tilde{y}^{2}+\diff\tilde{z}^{2}.
\]
Since here $K_{\mu\nu}=0$, one can see from the Ricci equation (\ref{Ricci})
that there is at least one component of the Riemann tensor proportional
to $\partial_{i}\partial_{j}N$ for each pair $\{i,j\}\in\{1,2,3\}^{2}$.
Requiring that $\mathcal{R}$ is flat therefore means that the most
general form that the lapse may take in this coordinate system is
linear in the spatial coordinates, giving 
\[
N(t,\tilde{\mathbf{x}})=p(t)+\mathbf{q}(t)\cdot\tilde{\mathbf{x}},
\]
for some $p$, $\mathbf{q}$ dependent only on $t$.

Looking at Figure \ref{Fig:=000020foliation}, it is now clear exactly
how rest frame transitions work geometrically. We simply have to choose
a geometry inside $\tilde{\mathcal{R}}$ such that the extension of
$\mathbf{n}$ inside $\tilde{\mathcal{R}}$ from the past and the
future do not agree, differing by a vector characterised by $\bm{\sigma}$.
Then free-falling observers will exit $\tilde{K}$ with a 3-velocity
of $-\bm{\sigma}$.

\section{Tables of symbols}\label{sec:Tables-of-symbols}

Due to the large number of functions and quantities represented by
different symbols throughout the paper, we provide a summary of the
most important ones in this appendix for ease of reference.

\subsection*{General}

The following symbols appear in several sections, with very similar
or identical meanings.
\begin{center}
\begin{tabular}{|>{\centering}m{1cm}|m{13cm}|}
\hline 
$u$ & Speed of destination rest frame as measured in passenger's initial
rest frame\tabularnewline
\hline 
$\gamma_{u}$ & Lorentz factor for speed $u$\tabularnewline
\hline 
$\Sigma_{t}$ & Hypersurface of constant $t$\tabularnewline
\hline 
$\mathbf{n},n^{\mu}$ & Future-pointing unit normal vector to $\Sigma_{t}$\tabularnewline
\hline 
$N$ & Lapse function associated to $t$ such that $\mathbf{n}=-N\diff t$\tabularnewline
\hline 
$\bm{\beta}$ & Shift vector\tabularnewline
\hline 
$\gamma_{ij}$ & Induced metric on hypersurface $\Sigma_{t}$\tabularnewline
\hline 
$\Gamma_{RFT}$ & Inextendible timelike geodesic corresponding to a rest frame transition\tabularnewline
\hline 
\end{tabular}
\par\end{center}

\subsection*{Section \ref{Sec:=000020Intro}: Introduction}
\begin{center}
\begin{tabular}{|>{\centering}m{1cm}|m{13cm}|}
\hline 
$\zeta(t)$ & Path in $z$ followed by Alcubierre drive\tabularnewline
\hline 
$v(t)$ & Velocity of Alcubierre drive, given by $\partial_{t}\zeta(t)$\tabularnewline
\hline 
$f$ & Shape function of Alcubierre drive\tabularnewline
\hline 
$\alpha(\tau)$ & Worldline of Alcubierre drive centre, a geodesic parameterised by
proper time\tabularnewline
\hline 
\end{tabular}
\par\end{center}

\subsection*{Section \ref{Sec:=000020Natario=000020failure}: Geodesic rest frame
transitions}
\begin{center}
\begin{tabular}{|>{\centering}m{1cm}|m{13cm}|}
\hline 
$T_{flat}$ & Time at which warp drive vanishes\tabularnewline
\hline 
$\mathbf{r}(t)$ & Path followed by warp drive centre\tabularnewline
\hline 
$\mathbf{v}(t,\mathbf{x})$ & Velocity vector field of warp drive; 3-velocity of warp drive's passenger
at $(t,\mathbf{r}(t))$\tabularnewline
\hline 
$s(t,\mathbf{x})$ & Function appearing in definition of lapse $N$\tabularnewline
\hline 
$T_{1}$, $\tau_{1}$ & Coordinate and proper time at which warp drive starts transition ($T_{1}=\tau_{1}$)\tabularnewline
\hline 
$T_{2}$, $\tau_{2}$ & Coordinate and proper time at which warp drive finishes transition
($T_{2}\neq\tau_{2}$)\tabularnewline
\hline 
$a(t)$ & Transition function for shift vector $\bm{\beta}$; decreases smoothly
from $1$ to $0$ in $T_{1}<t<T_{2}$\tabularnewline
\hline 
$b(t)$ & Transition function for lapse $N$; takes form of bump function between
$T_{1}<0<T_{2}$\tabularnewline
\hline 
$\lambda(t)$ & Function of $a$; analogous to Lorentz factor\tabularnewline
\hline 
$\chi_{1}(\tau)$ & Parametrisation of first part of geodesic inside warp drive by proper
time, before transition ($0\leq t<T_{1}$)\tabularnewline
\hline 
$\chi_{2}(\tau)$ & Parametrisation of second part of geodesic inside warp drive by proper
time, during transition ($T_{1}\leq t<T_{2}$)\tabularnewline
\hline 
$\chi(\tau)$ & Parametrisation of combination of $\chi_{1}$ and $\chi_{2}$ by proper
time; also a geodesic\tabularnewline
\hline 
\end{tabular}
\par\end{center}

\subsection*{Section \ref{Sec:=000020Making=000020CTC}: Creating a closed timelike
geodesic}

Every symbol describing the outgoing warp drive has a counterpart
with a hat. These are the corresponding symbols describing the returning
warp drive in the CTC spacetime, $M_{CTC}$. There are two other changes: 
\begin{enumerate}
\item The critical times, $t=0,T_{1},T_{2}$, now become $\hat{t}=\hat{T}_{0},\hat{T}_{1},\hat{T}_{2}$
respectively. 
\item $\hat{\mathbf{v}}$ finishes such that the warp drive is at rest in
$S$, so for $\hat{t}\geq\hat{T}_{1}$, $\hat{\mathbf{v}}=(0,0,-u)$. 
\end{enumerate}
\begin{center}
\begin{tabular}{|>{\centering}m{2cm}|m{12cm}|}
\hline 
$h_{\mu\nu},\hat{h}_{\mu\nu}$ & Perturbations to Minkowski metric in outgoing and returning warp drives'
domains respectively\tabularnewline
\hline 
$T_{finish}$ & Time when second warp drive returns, as measured in $S$\tabularnewline
\hline 
$\bar{v}^{z}$, $\hat{\bar{v}}^{z}$ & Average speeds (taken as positive) of both warp drives along their
respective $z$ axes in $S$ and $\hat{S}$ respectively\tabularnewline
\hline 
$C$ & Section of $t$ axis between $t=T_{finish}$ and $t=0$, assuming
$T_{finish}<0$\tabularnewline
\hline 
$\Gamma_{CTC}$ & Union of outgoing and returning geodesics at warp bubble centre and
$C$; itself a geodesic\tabularnewline
\hline 
$\epsilon,\hat{\epsilon}$ & Amount of proper time elapsed since start of transition\tabularnewline
\hline 
$\prescript{(CTC)}{}{\tensor{R}{_{\rho\sigma\mu\nu}}}$ & Riemann tensor on CTC spacetime\tabularnewline
\hline 
$\tensor{R}{_{\rho\sigma\mu\nu}},\tensor{\hat{R}}{_{\rho\sigma\mu\nu}}$ & Riemann tensors within respective domains of warp drives\tabularnewline
\hline 
$\gamma_{\mu\nu},\hat{\gamma}_{\mu\nu}$ & Projection operators for hypersurfaces $\Sigma_{t}$, $\hat{\Sigma}_{\hat{t}}$\tabularnewline
\hline 
$K_{\mu\nu},\hat{K}_{\mu\nu}$ & Extrinsic curvature tensors for hypersurfaces $\Sigma_{t}$, $\hat{\Sigma}_{\hat{t}}$\tabularnewline
\hline 
\end{tabular}
\par\end{center}

The specific example in subsection \ref{Sec:=000020specific=000020example}
adds the following symbols:
\begin{center}
\begin{tabular}{|>{\centering}m{1cm}|m{13cm}|}
\hline 
$q$ & Bump function with $0\leq q(x)\leq1$, $\supp{q}=(0,1)$ and $\int_{0}^{1}q(x)\diff x=1$;
here chosen as $q(x)=140x^{3}(1-x)^{3}$ for $0<x<1$\tabularnewline
\hline 
$q^{(-1)}$ & Primitive of $q$, a `smooth step function'; $q^{(-1)}(x)=0$ for
$x\leq0$, $q^{(-1)}(x)=1$ for $x\geq1$\tabularnewline
\hline 
$q^{(-2)}$ & Second primitive of $q$; $q^{(-2)}(x)=0$ for $x\leq0$\tabularnewline
\hline 
$r_{1}$ & Radius of flat region inside warp drives\tabularnewline
\hline 
$r_{2}$ & Radius of warp drives as measured by external observers in their respective
frames\tabularnewline
\hline 
$t_{1},\hat{t}_{1}$ & Times at which warp drives change from acceleration to deceleration\tabularnewline
\hline 
$t_{2}$, $\hat{t}_{2}$ & Actual finish times of transition, slightly less than $T_{2}$, $\hat{T}_{2}$,
to avoid collision of warp drives\tabularnewline
\hline 
$\alpha$ & Acceleration parameter\tabularnewline
\hline 
$\kappa$ & Constant associated with the choice of $q$; $\kappa=q^{(-2)}(1)$\tabularnewline
\hline 
$\omega$ & $T_{1}/T_{2}$, fraction of the outgoing journey spent before the
transition starts; same for both outgoing and returning sections\tabularnewline
\hline 
\end{tabular}
\par\end{center}

\subsection*{Section \ref{Sec:=000020energy=000020cond}: The weak energy condition
with a non-unit lapse}
\begin{center}
\begin{tabular}{|>{\centering}m{1cm}|m{13cm}|}
\hline 
$\rho$ & Eulerian energy density\tabularnewline
\hline 
$\phi_{i}$ & Momentum flux of energy-momentum tensor; $\phi_{i}=G_{ni}$\tabularnewline
\hline 
$\psi$ & Primitive of $\bm{\beta}$ in the curl-free case; $\bm{\beta}=\nabla\psi$\tabularnewline
\hline 
\end{tabular}
\par\end{center}

\subsection*{Appendix \ref{Sec:=000020FT=000020general=000020analysis}: General
analysis of rest frame transitions}
\begin{center}
\begin{tabular}{|>{\centering}m{1cm}|m{13cm}|}
\hline 
$M$ & Entire spacetime\tabularnewline
\hline 
$K$ & Compact region outside which $M$ is Riemann-flat\tabularnewline
\hline 
$F$ & Complement of $K$; Riemann-flat\tabularnewline
\hline 
$\mathcal{R}$ & Flat interior of warp bubble\tabularnewline
\hline 
$\Gamma$ & Inextendible timelike geodesic inside $\mathcal{R}$\tabularnewline
\hline 
$\tilde{\mathcal{R}}$ & Subset of $\mathcal{R}$ isometric to infinite 3+1 cylinder with $\Gamma\subset\tilde{\mathcal{R}}$\tabularnewline
\hline 
$\tilde{R}_{t}$ & Hypersurface inside $\tilde{\mathcal{R}}$ of time $t$\tabularnewline
\hline 
$T$ & Time at which observer in warp drive exits $\tilde{K}$\tabularnewline
\hline 
$\tilde{K}$ & Superset of $K$ such that $\tilde{\mathcal{R}}\cap\partial\tilde{K}$
is composed of two disjoint hypersurfaces, one at $t=0$ and one at
$t=T$\tabularnewline
\hline 
$\tilde{F}$ & Complement of $\tilde{K}$; Riemann-flat\tabularnewline
\hline 
$\mathbf{n}_{1}$ & Extension of $\mathbf{n}$ inside $\tilde{\mathcal{R}}$ from past
of $\tilde{K}\cap\tilde{\mathcal{R}}$\tabularnewline
\hline 
$\mathbf{n}_{2}$ & Extension of $\mathbf{n}$ inside $\tilde{\mathcal{R}}$ from future
of $\tilde{K}\cap\tilde{\mathcal{R}}$\tabularnewline
\hline 
$\bm{\sigma}$ & Proportional to spatial part of $\mathbf{n}_{2}$, when viewed in
$(\tau,x,y,z)$ coordinates\tabularnewline
\hline 
$\phi$ & Primitive of $\bm{\sigma}$, i.e. $\diff\phi=\bm{\sigma}$\tabularnewline
\hline 
$\tau_{0}$ & Infimum of $\tau$ on surface $t=T$\tabularnewline
\hline 
\end{tabular}
\par\end{center}

\bibliographystyle{unsrturl}
\phantomsection\addcontentsline{toc}{section}{\refname}\bibliography{WarpCTC}

\end{document}